\PassOptionsToPackage{unicode}{hyperref}
\PassOptionsToPackage{hyphens}{url}
\PassOptionsToPackage{dvipsnames,svgnames,x11names}{xcolor}
\documentclass[
  12pt]{article}

\usepackage{amsmath,amssymb}

\usepackage{caption}
\usepackage{graphicx}
\usepackage{dsfont}
\usepackage{bbm}
\usepackage{booktabs}
\usepackage{tabularx}
\usepackage{tikz}
\usetikzlibrary{arrows, positioning}

\usepackage{iftex}
\ifPDFTeX
  \usepackage[T1]{fontenc}
  \usepackage[utf8]{inputenc}
  \usepackage{textcomp} 
\else 
  \usepackage{unicode-math}
  \defaultfontfeatures{Scale=MatchLowercase}
  \defaultfontfeatures[\rmfamily]{Ligatures=TeX,Scale=1}
\fi
\usepackage{lmodern}
\ifPDFTeX\else  
\fi
\IfFileExists{upquote.sty}{\usepackage{upquote}}{}
\IfFileExists{microtype.sty}{
  \usepackage[]{microtype}
  \UseMicrotypeSet[protrusion]{basicmath} 
}{}
\makeatletter
\@ifundefined{KOMAClassName}{
  \IfFileExists{parskip.sty}{%
    \usepackage{parskip}
  }{
    \setlength{\parindent}{0pt}
    \setlength{\parskip}{6pt plus 2pt minus 1pt}}
}{
  \KOMAoptions{parskip=half}}
\makeatother
\usepackage{xcolor}
\makeatletter
\ifx\paragraph\undefined\else
  \let\oldparagraph\paragraph
  \renewcommand{\paragraph}{
    \@ifstar
      \xxxParagraphStar
      \xxxParagraphNoStar
  }
  \newcommand{\xxxParagraphStar}[1]{\oldparagraph*{#1}\mbox{}}
  \newcommand{\xxxParagraphNoStar}[1]{\oldparagraph{#1}\mbox{}}
\fi
\ifx\subparagraph\undefined\else
  \let\oldsubparagraph\subparagraph
  \renewcommand{\subparagraph}{
    \@ifstar
      \xxxSubParagraphStar
      \xxxSubParagraphNoStar
  }
  \newcommand{\xxxSubParagraphStar}[1]{\oldsubparagraph*{#1}\mbox{}}
  \newcommand{\xxxSubParagraphNoStar}[1]{\oldsubparagraph{#1}\mbox{}}
\fi
\makeatother

\usepackage{longtable,booktabs,array}
\usepackage{calc} 
\usepackage{etoolbox}
\makeatletter
\patchcmd\longtable{\par}{\if@noskipsec\mbox{}\fi\par}{}{}
\makeatother
\IfFileExists{footnotehyper.sty}{\usepackage{footnotehyper}}{\usepackage{footnote}}
\makesavenoteenv{longtable}
\usepackage{graphicx}
\makeatletter
\def\maxwidth{\ifdim\Gin@nat@width>\linewidth\linewidth\else\Gin@nat@width\fi}
\def\maxheight{\ifdim\Gin@nat@height>\textheight\textheight\else\Gin@nat@height\fi}
\makeatother
\setkeys{Gin}{width=\maxwidth,height=\maxheight,keepaspectratio}
\makeatletter
\def\fps@figure{htbp}
\makeatother

\makeatletter
\@ifpackageloaded{caption}{}{\usepackage{caption}}
\AtBeginDocument{%
\ifdefined\contentsname
  \renewcommand*\contentsname{Table of contents}
\else
  \newcommand\contentsname{Table of contents}
\fi
\ifdefined\listfigurename
  \renewcommand*\listfigurename{List of Figures}
\else
  \newcommand\listfigurename{List of Figures}
\fi
\ifdefined\listtablename
  \renewcommand*\listtablename{List of Tables}
\else
  \newcommand\listtablename{List of Tables}
\fi
\ifdefined\figurename
  \renewcommand*\figurename{Figure}
\else
  \newcommand\figurename{Figure}
\fi
\ifdefined\tablename
  \renewcommand*\tablename{Table}
\else
  \newcommand\tablename{Table}
\fi
}
\@ifpackageloaded{float}{}{\usepackage{float}}
\floatstyle{ruled}
\@ifundefined{c@chapter}{\newfloat{codelisting}{h}{lop}}{\newfloat{codelisting}{h}{lop}[chapter]}
\floatname{codelisting}{Listing}

\makeatother
\makeatletter
\@ifpackageloaded{caption}{}{\usepackage{caption}}
\@ifpackageloaded{subcaption}{}{\usepackage{subcaption}}
\makeatother

\ifLuaTeX
  \usepackage{selnolig}  
\fi
\usepackage[]{natbib}
\usepackage{bookmark}

\IfFileExists{xurl.sty}{\usepackage{xurl}}{} 
\hypersetup{
  pdftitle={Title},
  pdfauthor={Author 1; Author 2},
  pdfkeywords={3 to 6 keywords, that do not appear in the title},
  colorlinks=true,
  linkcolor={blue},
  filecolor={Maroon},
  citecolor={Blue},
  urlcolor={Blue},
  pdfcreator={LaTeX via pandoc}}

\newtheorem{prop}{Proposition}

\DeclareMathOperator{\expit}{expit}

\DeclareMathOperator{\Var}{Var}

\newcommand{\drmst}{\mathsf{dRMST}}

\newcommand\at[2]{\left.#1\right|_{#2}}

\long\def\ignore#1{}

\newcommand{\I}{\mathds{1}}
\newcommand{\Prob}{\mathsf{P}}
\newcommand{\Ex}{\mathsf{E}}
\newcommand{\Va}{V_{\tilde a}}
\newcommand{\va}{v_{\tilde a}}

\newcommand{\atil}{\tilde a}

\newtheorem{assumption}{Assumption}
\newtheorem{definition}{Definition}
\newtheorem{theorem}{Theorem}

\newcommand{\anon}{1}

\begin{document}

\def\spacingset#1{\renewcommand{\baselinestretch}%
{#1}\small\normalsize} \spacingset{1}


\if1\anon
{
  \title{\bf Causal Survival Analysis in Platform Trials with Non-Concurrent Controls}
  \author{Antonio D'Alessandro\thanks{
    \textit{This article is based upon work supported by the National Science Foundation under Grant No 2306556,  and the National Institute of Health Grant No 1R01AI197146-01}}\hspace{.2cm}\\
    and \\
    Samrachana Adhikari \\
    and \\
    Michele Santacatterina \\
    Division of Biostatistics, Department of Population Health, New York University
    }
  \maketitle
} \fi

\if0\anon
{
  \bigskip
  \bigskip
  \bigskip
  \begin{center}
    {\LARGE\bf Title}
\end{center}
  \medskip
} \fi

\bigskip
\begin{abstract}
Platform trials allow treatment arms to enter and exit over time while maintaining a shared control arm, yielding concurrent and non-concurrent controls (NCC). Pooling NCC is often motivated as a strategy to improve statistical efficiency, but it is unclear which estimand is targeted, what assumptions justify identification and estimation, and when precision gains are achievable; these questions are further complicated by time-to-event outcomes. Motivated by the Adaptive COVID-19 Treatment Trial (ACTT), with time to recovery as the primary endpoint, we develop an estimand-first causal survival framework targeting the treatment-specific counterfactual survival curve in the concurrently eligible population and functionals such as the restricted mean survival time (RMST). We provide nonparametric identification results and formalize the pooling assumption required for NCC to identify the concurrent control hazard. We study covariate-adjusted outcome-regression (OR) and doubly robust (DR) estimators, comparing concurrent-only and pooled-control versions. Pooling can improve precision for OR estimators when the pooling assumption holds and parametric hazard models are correctly specified, but can induce bias and invalidate inference under misspecification. For DR estimators, pooling NCC gives no first-order efficiency gain when treatment availability is deterministic in entry time; possible gains under stochastic availability still rely on pooling assumptions. These results support targeting the concurrent causal survival estimand, avoiding pooling unless the required assumptions are credible, and using covariate-adjusted DR estimation with concurrent controls only. Simulations and an ACTT application illustrate the framework.
\end{abstract}

\noindent%
{\it Keywords:}  Causal survival estimands; Doubly robust estimation; Restricted mean survival time
\vfill

\newpage
\spacingset{1.8} 

\section{Introduction}\label{sec:intro}

A platform trial is an adaptive multi-arm experimental design that allows for the simultaneous evaluation of multiple treatments for a single disease \citep{berry2015}. Traditional randomized controlled trials (RCTs) feature parallel arms which remain fixed by design through the duration of the experiment, while platform designs allow different treatment arms to enter and exit at distinct times, while also maintaining a shared control arm through the end of the study.  This shared control arm is composed of \textit{concurrent controls}, subjects who entered the trial concurrent with the treatment of interest and could have been randomized to the treatment of interest or control arm, and \textit{non-concurrent controls} (NCC), subjects who entered the trial when the treatment of interest was unavailable and had zero probability of being assigned to the treatment of interest. This design has seen widespread use during the COVID-19 pandemic \citep{hayward2021}. For example, the Adaptive COVID-19 Treatment Trial (ACTT) \citep{kalil2021} evaluated multiple regimens across stages, with remdesivir serving as a shared arm across ACTT-1 and ACTT-2.

Time-to-event endpoints are common in platform trials and require methods that accommodate censoring. Although analyses often rely on the Cox proportional hazards model \citep{cox1972}, the hazard ratio targets an instantaneous risk-set contrast that may not correspond to a clinically meaningful causal effect on survival time \citep{hernan2010hazards, stensrud2019limitations}. This motivates complementary estimands such as the restricted mean survival time (RMST), which summarizes average survival up to a clinically relevant horizon and yields interpretable contrasts between treatment arms \citep{royston2013}.

While platform trials like the ACTT offer increased flexibility, they also create an apparent opportunity to increase precision by pooling earlier-stage control patients with concurrent controls for a given treatment comparison. However, it is not generally clear when such pooling is valid, what causal quantity is targeted, or how pooling should be implemented for time-to-event endpoints in the presence of censoring. This raises three \textit{practical} questions: \textit{1) how should causal survival estimands be defined when controls are partly nonconcurrent; 2) under what assumptions are these estimands identifiable and estimable; and 3) when, if at all, does incorporating NCC improve precision for the concurrent treatment comparison?} Answering these questions requires thinking carefully about how calendar time can enter the data-generating process in a platform trial. When the distribution of baseline covariates change over the enrollment period, na\"{\i}vely pooling NCC and concurrent controls can bias estimated treatment effects; this is commonly referred to as ``time drift'' or ``temporal drift''. Proposed solutions to deal with this bias include test-then-pool procedures \citep{viele2014}, regression adjustment \citep{Saville2022}, propensity weighting \citep{Chen2020}, and robust borrowing approaches \citep{ kaizer2018, wang2023}.

While these approaches enable the use of NCC, the causal target is often implicit and the estimand can be tied closely to a chosen modeling strategy (e.g., a particular regression coefficient or hazard ratio) rather than stated upfront as a clinically meaningful causal contrast and research question. This emphasis on specifying the scientific question first is central to the estimand-first framework advocated by the FDA \citep{food2021e9} and the ICH E9(R1) guidance \citep{ich2017}. Recent reviews note that, despite substantial methodological development, there remains a need for a unified estimand-first framework for platform trials that clearly links causal questions, identifying assumptions, and estimation strategies, including for time-to-event endpoints \citep{collignon2022estimands, koenig2024current}.

\subsection{Our contributions and recommendations}

Motivated by the ACTT platform trial with time to recovery as the primary endpoint, this paper addresses these questions using an estimand-first causal survival framework. We target the treatment-specific counterfactual survival curve in the concurrently eligible population and functionals such as the RMST contrast. This framing separates three issues that are often conflated in analyses using non-concurrent controls: the causal estimand, the pooling assumption needed to use non-concurrent controls, and the source of any precision gain.

We show that borrowing non-concurrent controls is not justified by the platform structure alone. It requires a pooling assumption under which the control hazard learned from pooled controls represents the control hazard in the concurrent population after adjustment for entry time and baseline covariates. This assumption is formalized in Assumption~\ref{ass:pool}. When treatment availability is deterministic in entry time, Assumption~\ref{ass:pool} holds on the concurrent support. When treatment availability is stochastic, Assumption~\ref{ass:pool} becomes a substantive pooling assumption.

We then study parametric outcome-regression and covariate-adjusted doubly robust estimators in Sections~\ref{sec:estor} and~\ref{sec:estdr}, leading to a cautionary conclusion. Pooled-control outcome-regression estimators can improve precision only when Assumption~\ref{ass:pool} holds and the parametric hazard model is correctly specified; otherwise, pooling can induce bias (see Section~\ref{sec:efficiency} and Section~\ref{sec:appendix_A7} in the Supplementary Material). For doubly robust estimators, pooling non-concurrent controls gives no efficiency gain when treatment availability is deterministic in entry time, while possible gains under stochastic treatment availability still rely on Assumption~\ref{ass:pool} (see Section~\ref{sec:efficiency}). These results support the practical recommendation: target the concurrent causal survival estimand, avoid pooling unless the required assumptions are credible, and use covariate-adjusted doubly robust estimation with concurrent controls only. The simulation study in Section~\ref{sec:sim} and the ACTT application in Section~\ref{sec:rwd} illustrate the proposed framework.

\section{Additional Related Work}
\label{sec:work}
Platform trials extend traditional parallel-arm designs through features such as shared control arms and adaptive treatment entry and removal \citep{berry2016, aptc2019}. Methodological development continues from both Bayesian and frequentist perspectives \citep{mu2024, greenstreet2025}, with recent reviews documenting the diversity of design characteristics and analytic strategies \citep{masson2025}. A central statistical question concerns how best to leverage shared or pooled control information. Borrowing of control data, including from external sources, has a long history \citep{viele2014}, and its potential efficiency gains in platform settings have been studied in  \cite{huang2023} and \cite{santa2025}; see \cite{roig2023} for a review.

For time-to-event endpoints, both frequentist and Bayesian approaches exploit shared controls. Frequentist proposals build on multi-arm multi-stage survival designs using standard time-to-event methods under proportional hazards working models \citep{royston2011designs}. Bayesian approaches include adaptive platform designs such as MIDAS \citep{yuan2016} and commensurate or MAP-style borrowing strategies for censored outcomes \citep{roychoudhury2020bayesian}. Applications include REMAP-CAP \citep{remap2021} and the HEALEY ALS Platform Trial \citep{healey2025}. Across these strands, inference is typically framed in terms of model-based parameters (e.g., hazard ratios or borrowing parameters), underscoring the importance of clearly specifying the causal estimand. We focus on estimating causal survival contrasts in platform trials using the restricted mean survival time (RMST), which offers interpretational advantages over hazard ratios \citep{hasegawa2020}; see \cite{kloecker2020} for applied illustrations. Our estimators rely on semiparametric efficiency theory and double robustness \citep{kennedy2021semiparametric, hines2021}, building on weighted estimating-equation approaches originally developed in \cite{robins1994}. 

Conceptually, the use of non-concurrent controls can be viewed as a transportability problem: information from controls enrolled under one availability regime is used to learn about the concurrent control outcome process only under explicit assumptions linking the relevant hazard functions across regimes \citep[among others]{pearl2011transportability,degtiar2023review,rudolph2022efficiently}.

\section{Notation and setup}
\label{sec:note}

Suppose we observe an independent and identically distributed (iid) sample $(Z_1,\ldots,Z_n)$ drawn from some joint distribution $\mathsf{P}$, assumed to be an element of a nonparametric model. For each $i\in\{1,\ldots,n\}$, let $W_i$ denote baseline covariates and let $E_i$ denote the (random) study entry time and $X_i = (W_i, E_i)$ be the combined vector. Let $A_i$ denote the randomized treatment assignment, taking values in $\{0,\ldots,\mathcal{A}\}$, where $A_i=0$ denotes the shared control arm and $A_i=a$ for $a\in\{1,\ldots,\mathcal{A}\}$ denotes active treatment $a$.  Throughout, we use $a\in\{0,\ldots,\mathcal{A}\}$ as a generic arm index (with $a=0$ for control). 
Let $V_i=(V_{0,i},\ldots,V_{\mathcal{A},i})$ denote the treatment availability vector at entry, where $V_{0,i}\equiv 1$ and, for $a\in\{1,\ldots,\mathcal A\}$, $V_{a,i}\in\{0,1\}$ indicates whether treatment $a$ was available to subject $i$ at entry time $E_i$. Availability may be deterministic in $E_i$ or stochastic conditional on $E_i$ through the trial design. We also fix an active treatment index $\tilde a\in\{1,\ldots,\mathcal{A}\}$ to define the concurrent population for the comparison of treatment $\tilde a$ versus control.
The observed data are $D=(Z_1,\ldots,Z_n)$ with
$Z_i=(W_i,E_i,A_i,V_i,\Delta_i,\tilde{T}_i)\sim \mathsf{P}$.
For notational simplicity, we omit the subject index $i$ when no confusion can arise and write $V_a$ for the $a$th component of $V$. Note that $\mathsf{P}(A=\tilde a \mid V_{\tilde a}=0)=0$ by design. Further, assume $K$ equally spaced discrete time points $\{1,\ldots,K\}$. 
Let $T$ be the time-to-event outcome variable taking values in $\{1, \ldots,K\} \hspace{1mm} \cup \{\infty\}$ where $T = \infty$ corresponds to no event occurring during any time in $1,\ldots ,K$. 
Let $C \in \{0,\ldots,K\}$ represent the censoring time defined to be the time at which the subject was last observed in the study. If a subject remains through to the end of the study we say $C=K$ and refer to this as administrative censoring. We assume no censoring occurs before follow-up begins, so \(C=0\) is not observed in the analytic data. Let the observed event time be denoted by $\tilde{T} = \text{min}(C,T)$ and let $\Delta= \I(T\leq C)$ be the indicator variable recording whether a subjects event time was observed ($\Delta = 1$) or censored ($\Delta =0$). 
For each subject, the observed vector $Z$ may be equivalently represented using a longitudinal data structure given by:
$Z = (W,E,A,V,R_0,L_1,R_1,L_2,\ldots,R_{K-1},L_K)$where for $t=0,\ldots,K-1$ we define $R_t = \I(\tilde{T}=t, \Delta=0)$ and for $t=1,\ldots,K$ we define $L_t=\I(\tilde{T} =t,\Delta=1)$. Because baseline censoring is excluded from the analytic data, \(R_0\equiv 0\); it is retained only to keep the longitudinal representation and history notation uniform. The sequence of indicators
($R_0,L_1,R_1,L_2,\ldots,R_{K-1},L_{K}$) is composed of all zeros until the first time the event is observed or censoring occurs (and if all indicators are zero, we set $C=K$, $\Delta=0$, and $\tilde{T}=K$, i.e., the subject is observed event-free through time $K$).
Additionally, for any random variable $H$ we define the history of $H$ through time $t$ as $\bar{H}_t = (H_0,\ldots,H_t)$. For a given scalar $h$, the expression $\bar{H}_t=h$ denotes element-wise equality, and we adopt the convention $L_0\equiv 0$. Using this idea of a random variable's history we introduce two additional indicator functions
$I_t  = \I(\bar{R}_{t-1}=0,\bar{L}_{t-1}=0)$ and $J_t = \I(\bar{R}_{t-1}=0,\bar{L}_{t}=0)$ for each $t \geq 1$. The variable $I_t$ denotes whether a subject is at risk of the event being observed at time $t$, analogously $J_t$  denotes whether a subject is at risk of censoring at time $t$ where we let $J_0=1$ by convention.

\subsection{Time-to-event quantities}

In this section we introduce statistical quantities used in the analysis of time-to-event outcomes. Using the previously established notation, fix $\tilde a\in\{1,\ldots,\mathcal A\}$ and consider the two-arm comparison $a\in\{0,\tilde a\}$ within the concurrent population $\{V_{\tilde a}=1\}$.

The discrete-time hazard for the event at $m\in\{1,\ldots,K\}$ is
\[
\begin{aligned}
h(m,a,\tilde a,e,w)
&= \Prob(T = m \mid T \ge m, C \ge m, A=a, V_{\tilde a}=1, E=e, W=w) \\
&= \Prob(L_m=1 \mid I_m=1, A=a, V_{\tilde a}=1, E=e, W=w).
\end{aligned}
\]

The discrete-time censoring hazard for $m\in\{1,\ldots,K-1\}$ is
\[
\begin{aligned}
g(m,a,\tilde a,e,w)
&= \Prob(C = m \mid T > m, C \ge m, A=a, V_{\tilde a}=1, E=e, W=w) \\
&= \Prob(R_m=1 \mid J_m=1, A=a, V_{\tilde a}=1, E=e, W=w).
\end{aligned}
\]

Fix a truncation time $\tau\in\{1,\ldots,K\}$. For $t\in\{0,\ldots,\tau\}$ and $a\in\{0,\tilde a\}$ define
\[
\begin{aligned}
S(t\mid A=a,V_{\tilde a}=1,E=e,W=w)
&= \Prob(T>t\mid A=a,V_{\tilde a}=1,E=e,W=w) \\
G(t\mid A=a,V_{\tilde a}=1,E=e,W=w)
&= \Prob(C\ge t\mid A=a,V_{\tilde a}=1,E=e,W=w).
\end{aligned}
\]

Finally, denote the propensity function by, 
\[
\begin{aligned}
\pi(a,\tilde a,w,e)=\Prob(A=a\mid V_{\tilde a}=1,E=e,W=w),\qquad a\in\{0,\tilde a\}.
\end{aligned}
\]

\subsection{Causal model and associated directed acyclic graph}

To formalize entry time, covariates, treatment assignment, and treatment availability in a platform trial, we adopt the following structural causal model \citep{pearl1995causal},
\begin{align}\label{eq:npsem}
    E = f_E(U_E), \quad W = f_W(E, U_W),\quad\quad\quad\quad\quad\quad\quad\\ \notag
    \quad V = (V_0, \ldots V_{\mathcal{A}}),  \quad  V_{a} = f_{V_{a}}(E, U_{V_{a}}) \quad a \in \{0,\ldots,\mathcal{A}\},\quad\\
    A = f_A(W,V,U_A), \quad T = f_T(A,W,E, U_T), \quad C = f_C(A, W, E, U_C). \notag
\end{align}

Fix $\tilde a \in {1,\ldots,\mathcal A}$ and consider the comparison of treatment $\tilde a$ versus shared control ($A=0$) within the concurrent population ${V_{\tilde a}=1}$, assuming the control arm is always available ($V_0\equiv 1$).
The model allows variables to depend on entry time $E$, accommodating temporal drift. Treatment assignment $A$ may depend on baseline covariates $W$ (e.g., stratified randomization) but depends on $E$ only through treatment availability $V$, so subjects are randomized only among arms available at entry, with fixed randomization probabilities over time.
Event and censoring times $(T,C)$ depend on realized treatment $A$ and entry time $E$, but not directly on availability $V$, and no direct causal relationship between $T$ and $C$ is imposed. Availability $V$ depends on calendar time but not on individual covariates $W$. Baseline covariates $W$ may vary with entry time, while $E$ does not depend on $W$, allowing population composition to evolve over time.
Functional forms for $f_E, f_W, f_T,$ and $f_C$ are unrestricted, whereas treatment assignment and availability are known by design and may be deterministic or stochastic in entry time. Exogenous variables $U_E, U_W, U_T, U_{V_a}, U_C$ represent unmeasured factors, and $U_A$ encodes the randomization mechanism.

\section{Definition and identification of the concurrent treatment-specific counterfactual survival curve}
\label{sec:ident}

Throughout this paper, we employ counterfactual notation \citep{pearl2010introduction}. For each
$a\in\{0,\tilde a\}$, let $T(a,K)$ denote the event time that would be observed under the joint intervention
setting treatment assignment to $A=a$ and eliminating censoring by setting the censoring time to $C=K$
with probability one. For notational simplicity, we write $T(a)\equiv T(a,K)$ throughout. For a cutoff denoted by $\tau \in \{1,\ldots,K\}$ the target estimand of interest is the area under the treatment specific survival curve among those patients in the concurrent population:

We focus on the comparison of the active treatment $\tilde a$ versus control within the
\emph{concurrent population} for $\tilde a$, i.e., among those with $V_{\tilde a}=1$.

\begin{definition}[Concurrent treatment-specific counterfactual survival curve]
\label{def:curve}
For $a\in\{0,\tilde a\}$ and $t\in\{0,1,\ldots,K\}$, the concurrent treatment-specific
counterfactual survival curve is
\[
\theta(a,t) = S_a^\ast(t)
:= \Prob\{T(a) > t \mid V_{\tilde a}=1\}.
\]
\end{definition}

The curve $\theta(a,t)$ is defined in terms of counterfactual outcomes $T(a)$ and is therefore not, in general,
a functional of the observed-data distribution without additional assumptions. In particular, for any given
subject we observe at most one of the potential event times $\{T(0),T(\tilde a)\}$, so the joint distribution of
counterfactual outcomes is not identified from the observed data alone. Nonparametric identification provides
conditions under which $S_a^\ast(t)$ can nonetheless be expressed uniquely as a functional of the observed-data
distribution $\mathsf P$, without imposing parametric restrictions on outcome or censoring models
\citep{pearl1995causal}. Such an observed-data representation motivates estimators and frequentist inference.
To discuss identification of $\theta(a,t)$, we introduce the following assumptions:

\begin{assumption}[Exchangeability]\label{ass:exchange} For each $a\in\{0,\tilde a\}$,
\[
T(a) \perp A \ \big|\ (W,E,V_{\tilde a}=1).
\]
\end{assumption}
Assumption \ref{ass:exchange}  states that within the concurrent population and conditional on $(W,E)$, treatment is as-if randomized so there is no residual confounding of $A$ and $T(a)$; this is fundamentally untestable because it involves counterfactual outcomes, however we expect it to hold by design under proper randomization.

\begin{assumption}[Consistency]\label{ass:cons} If $A=a$, then the latent event time satisfies $T = T(a)$ for $a\in\{0,\tilde a\}$.
\end{assumption}
Assumption \ref{ass:cons} states that for subjects who in fact receive arm $a$, the observed event time equals the corresponding potential event time $T(a)$; this is a definitional/measurement assumption (not empirically testable in the counterfactual sense) and can fail only via treatment-version ambiguity or interference.

\begin{assumption}[Random censoring]\label{ass:cens}For each $a\in\{0,\tilde a\}$,
\[
T(a) \perp C \ \big|\ (A,E,W,V_{\tilde a}=1).
\]
\end{assumption}
Given $(A,E,W)$ in the concurrent population, censoring carries no additional information about $T(a)$ beyond these variables; this is untestable without additional structure.

\begin{assumption}[Positivity of shared control assignment]\label{ass:pshare} For all $(e,w)$ such that $\Prob(E=e,W=w)>0$,
\[
\Prob(A=0 \mid E=e, W=w) > 0.
\]
\end{assumption}
Every $(E,W)$ stratum that occurs in the data has nonzero probability of being randomized to control, preventing empty control strata; this is checkable from the design and the realized data (violations appear as zero counts/zero estimated probabilities). This assumption should hold by design.
Note that, because the shared control arm is available throughout the platform trial ($V_0\equiv 1$), the positivity condition for control assignment does not depend on treatment availability; consequently, we state Assumption~\ref{ass:pshare} unconditionally in $(E,W)$ (and it therefore holds, in particular, within the concurrent population $\{V_{\tilde a}=1\}$).

\begin{assumption}[Probability of censoring]\label{ass:pcens} For all $t\in\{0,\ldots,K-1\}$,
\[
\Prob(C=t \mid C\ge t,\ A=a,\ E=e,\ W=w,\ V_{\tilde a}=1) < 1
\quad \text{for each } a\in\{0,\tilde a\}.
\]
\end{assumption}
Among those still under follow-up, censoring is not deterministic at any time point within any $(A,E,W)$ stratum in the concurrent population; this assumption can be checked empirically (violations appear as $\Prob(C=t\mid C\ge t,\cdot)=1$ in some strata/time points).

\begin{assumption}[Positivity among concurrent units]
\label{ass:pconc}
For all $(e,w)$ such that $\Prob(E=e,W=w,V_{\tilde a}=1)>0$,
\[
\Prob(A=\tilde a \mid E=e, W=w, V_{\tilde a}=1) > 0.
\]
\end{assumption}
Within the concurrent population $\{V_{\tilde a}=1\}$, every $(E,W)$ stratum that occurs has nonzero probability of assignment to $\tilde a$, ensuring overlap for the $\tilde a$ versus control comparison; this is primarily a design/implementation condition (checkable from the randomization scheme), though finite-sample data can still exhibit practical violations (e.g., zero counts in some strata).

\begin{assumption}[Pooling concurrent and nonconcurrent controls for the hazard]\label{ass:pool}
For each $m\in\{1,\ldots,K\}$ and all $(e,w)$ such that
\[
\Prob(I_m=1,A=0,V_{\tilde a}=1,E=e,W=w)>0
\quad\text{and}\quad
\Prob(I_m=1,A=0,E=e,W=w)>0,
\]
\[
h(m,0,\tilde a,e,w)
=
h(m,0,e,w),
\]
\end{assumption}
\noindent
where 
\begin{align*}
    h(m,0,e,w) &= \Prob(T = m \mid T \ge m,\; C \ge m,\; A = 0,\; E=e,\; W=w) \notag\\
    &= \Prob(L_m=1\mid I_m=1,\;A=0,\;E=e,\;W=w).
\end{align*}

Assumption~\ref{ass:pool} states that, after conditioning on $(E,W)$ (and being at risk at $m$), the control-arm event hazard is the same in the concurrent subset and in the pooled (concurrent plus nonconcurrent) controls, so pooled control data can be used to learn the concurrent control hazard.  Section~\ref{sec:appendix_A7} in the Supplementary Material provides additional details on Assumption~\ref{ass:pool}, including its role in identification, conditions under which it is plausible, and how it interacts with parametric model specification. 

Finally, the identification of the causal parameter given in Definition \ref{def:curve} requires the following preliminary results,

\begin{prop}[Product representation] \label{prop1}
 Under assumptions \ref{ass:exchange} - \ref{ass:pshare} the distribution functions for survival and censoring admit the following product representation: \\
\begin{equation*}
    S(t\mid A=a,V_{\tilde a}=1,E=e,W=w)
=\prod_{m=1}^{t}\{1-h(m,a,\tilde a,e,w)\}, \\
\end{equation*}
\begin{equation*}
    G(t\mid A=a,V_{\tilde a}=1,E=e,W=w)
=\prod_{m=1}^{t-1}\{1-g(m,a,\tilde a,e,w)\} \\
\end{equation*}
\end{prop}

Using these propositions it is possible to link the concurrent treatment-specific counterfactual survival curve to the observed data distribution,

\begin{theorem}[Nonparametric identification of $\theta(a,t)$]
\label{theorem1}
Fix $\tilde a\in\{1,\ldots,\mathcal A\}$ and let $a\in\{0,\tilde a\}$. Assume Model~\ref{eq:npsem},
Assumptions~\ref{ass:exchange}, \ref{ass:cons}, \ref{ass:cens}, \ref{ass:pshare}, \ref{ass:pcens}, \ref{ass:pconc},
and Proposition~\ref{prop1}. Then for each $t\in\{0,\ldots,\tau\}$,
\[
\theta(a,t)
:= \Prob\{T(a)>t \mid V_{\tilde a}=1\}
=
\Ex\!\left[\prod_{m=1}^{t}\{1-h(m,a,\tilde a,E,W)\}\ \big|\ V_{\tilde a}=1\right].
\]
In particular, for $a=0$,
\[
\theta(0,t)
=
\Ex\!\left[\prod_{m=1}^{t}\{1-h(m,0,\tilde a,E,W)\}\ \big|\ V_{\tilde a}=1\right].
\]
If, in addition, Assumption~\ref{ass:pool} holds, then $h(m,0,\tilde a,E,W)=h(m,0,E,W)$ for $V_{\tilde a}=1$, therefore
\[
\theta(0,t)
=
\Ex\!\left[\prod_{m=1}^{t}\{1-h(m,0,E,W)\}\ \big|\ V_{\tilde a}=1\right],
\]
which motivates estimating the control hazard using pooled concurrent and nonconcurrent controls.
\end{theorem}

The proof of Proposition~\ref{prop1} and Theorem~\ref{theorem1} can be found in the Supplementary Material. The first identification result in Theorem~\ref{theorem1}, which expresses the concurrent treatment-specific survival curve using the concurrent arm-specific hazard, follows standard identification arguments for treatment-specific survival curves under random censoring \citep[e.g.,][]{west2024}. The additional contribution here is the pooled-control representation for \(a=0\), which shows that non-concurrent controls can be used to learn the concurrent control component only when Assumption~\ref{ass:pool} justifies replacing \(h(m,0,\tilde a,E,W)\) by \(h(m,0,E,W)\).

\subsection{Restricted mean survival time}

The restricted mean survival time (RMST) is an alternative measure of time-to-event quantities in survival analysis.  The RMST represents the area under the survival curve from time 0 to a given cut-off, in this work denoted $\tau$. The interpretation is straight forward, the RMST is the  average time-to-event during the time period between 0 and $\tau$ \citep{hasegawa2020}. Contrasts such as the difference in restricted mean survival time thus represent the difference in the average time-to-event between two specified groups such as treatment vs. control or new treatment vs. standard of care. In addition to the easy interpretation, the RMST and related contrasts remain valid even when the proportional hazards assumption does not hold \citep{royston2013}, a common but often tenuous assumption in medical studies.  
Given the advantages of the RMST (and its contrasts), the causal estimand of interest in this work will be the difference in restricted mean survival time in the concurrently eligible population.
The identification result in Theorem \ref{theorem1} allows functions of the treatment-specific survival curve, like the RMST to be similarly identified, 

\begin{definition}[Difference in restricted mean survival time ($\drmst$)]\label{def:drmst}
Using Theorem~\ref{theorem1}, for $\tilde a \in \{1,\ldots,\mathcal{A}\}$ and $\tau \in \{1,\ldots,K\}$,
\begin{align*}
    \drmst(\tilde a,\tau)
    &= \Ex[ \min(T(\tilde a), \tau) - \min(T(0), \tau)  \mid V_{\tilde a}=1] \\
    &= \sum_{t=1}^{\tau-1} \{S^*_{\tilde a}(t) - S^*_0(t)\} \\
    &= \sum_{t=1}^{\tau-1} \{\theta(\tilde a,t)-\theta(0,t)\} \\
    &= \sum_{t=1}^{\tau-1}
    \Bigg\{
    \Ex\Big[ \prod_{m=1}^{t} \{1-h(m,\tilde a,\tilde a,E,W)\} \mid V_{\tilde a}=1\Big]
    -
    \Ex\Big[ \prod_{m=1}^{t} \{1-h(m,0,\tilde a,E,W)\} \mid V_{\tilde a}=1\Big]
    \Bigg\}.
\end{align*}
\end{definition}

In the following section we will discuss different estimators for $\drmst$ based on outcome regression and influence functions.

\subsection{Other survival contrasts}
\label{sec:other_surv_contrasts}

In this paper we focus on the difference in restricted mean survival time, which, as shown in Definition~\ref{def:drmst}, is a linear functional of \(\theta(a,t)\). Other contrasts can also be constructed from the concurrent treatment-specific survival curve depending on the scientific question \citep{west2024}. For example, when the event is recovery, as in the ACTT application \citep{kalil2021},  the \(t\)-day recovery ratio $\frac{\theta(0,t)}{\theta(\tilde a,t)},$ which compares the cumulative probability of recovery by time \(t\) under treatment \(\tilde a\) versus control. Other possible contrasts include the survival ratio \(\theta(\tilde a,t)/\theta(0,t)\), the risk difference \(\{1-\theta(\tilde a,t)\}-\{1-\theta(0,t)\}\) and the risk ratio \(1-\theta(\tilde a,t)/1-\theta(0,t)\).

\section{Estimation}
\label{sec:est}

For notational simplicity in the estimation section, we write the treatment of interest as $A=1$, corresponding to $A=\tilde a$, and the shared control as $A=0$. We will begin our discussion of estimation by introducing outcome regression (OR) estimators to help build intuition. However, outcome regression estimators are subject to model misspecification, i.e. if the functional form of the models are incorrect bias will be introduced. To address this limitation we then introduce doubly robust (DR) estimators. 

\subsection{Estimators based on parametric outcome regression} \label{sec:estor}

Using Theorem~\ref{theorem1}, an outcome-regression estimator using only concurrent controls can be defined as
\begin{align*}
\widehat{\drmst}_{OR}^{oc}
&=
\frac{\sum_{i=1}^{n}\sum_{t=1}^{\tau-1}
\widehat{\mathbf S}_{1}(i,t)\I(V_{\tilde a,i}=1)}
{\sum_{i=1}^{n}\I(V_{\tilde a,i}=1)}
-
\frac{\sum_{i=1}^{n}\sum_{t=1}^{\tau-1}
\widehat{\mathbf S}^{oc}_{0}(i,t)\I(V_{\tilde a,i}=1)}
{\sum_{i=1}^{n}\I(V_{\tilde a,i}=1)} .
\end{align*}

Alternatively, using the pooled-control representation in Theorem~\ref{theorem1} under Assumption~\ref{ass:pool}, an outcome-regression estimator using both concurrent and non-concurrent controls can be defined as
\begin{align*}
\widehat{\drmst}_{OR}^{ac}
&=
\frac{\sum_{i=1}^{n}\sum_{t=1}^{\tau-1}
\widehat{\mathbf S}_{1}(i,t)\I(V_{\tilde a,i}=1)}
{\sum_{i=1}^{n}\I(V_{\tilde a,i}=1)}
-
\frac{\sum_{i=1}^{n}\sum_{t=1}^{\tau-1}
\widehat{\mathbf S}^{ac}_{0}(i,t)\I(V_{\tilde a,i}=1)}
{\sum_{i=1}^{n}\I(V_{\tilde a,i}=1)} .
\end{align*}

Here $\widehat{\mathbf S}_{1}(i,t)=\prod_{m=1}^{t}\{1-\hat h(m,A=1,V_{\tilde a}=1,E_i,W_i)\}$ is obtained from the fitted active-arm hazard among concurrent participants. For the control arm, the concurrent-only estimator uses $\widehat{\mathbf S}^{oc}_{0}(i,t)=\prod_{m=1}^{t}\{1-\hat h^{oc}(m,A=0,V_{\tilde a}=1,E_i,W_i)\}$, where $\hat h^{oc}$ is learned using concurrent controls only. The all-controls estimator uses $\widehat{\mathbf S}^{ac}_{0}(i,t)=\prod_{m=1}^{t}\{1-\hat h^{ac}(m,A=0,E_i,W_i)\}$, where $\hat h^{ac}$ is learned using the full shared control arm. Thus, the difference between $\widehat{\drmst}_{OR}^{oc}$ and $\widehat{\drmst}_{OR}^{ac}$ lies only in the construction of the estimated control survival curve.
Under regularity conditions and correct specification of the fitted hazard models, $\widehat{\drmst}_{OR}^{oc}$ is consistent and asymptotically normal \citep{boos2013essential} for the concurrent estimand. The all-controls estimator $\widehat{\drmst}_{OR}^{ac}$ additionally requires Assumption~\ref{ass:pool}, so that the pooled control hazard targets the concurrent control hazard. Standard error estimates were computed using a modified bootstrap designed to sample from the longitudinal transformation of the survival data. Details for how to construct both OR estimators can be found in the Supplementary Material(Section \ref{sec:appendix_OR}).

\subsection{Doubly robust estimators} \label{sec:estdr}

Doubly robust (DR) estimators combine outcome regression and inverse-probability weighting. In the present setting, they are useful because consistency can be retained if the outcome-regression nuisance functions are consistently estimated or if the censoring and treatment-assignment nuisance functions are consistently estimated. We construct DR estimators by deriving influence-function-based estimating equations, following \cite{bickel1993efficient}, \cite{hines2021}, and \cite{kennedy2021semiparametric}. Influence functions provide a systematic way to construct estimators with asymptotic linearity and valid large-sample inference. They also allow flexible or data-adaptive nuisance-function estimation while retaining standard inference under suitable regularity conditions \cite{kennedy2021semiparametric}. For the all-controls DR estimator, pooling the control-arm censoring nuisance function requires the censoring analogue of Assumption~\ref{ass:pool}. Specifically, for each $m$ and all relevant $(e,w)$, 
\[ 
g(m,0,\tilde a,e,w)=g(m,0,e,w). 
\] 
This condition is not needed for identification of the concurrent survival curve in Theorem~\ref{theorem1}, but is needed below because $Q_0^{ac}$ below uses the pooled censoring survival $G(\cdot\mid A=0,X)$.

The next result gives the influence functions for the fixed-time contrast \(\theta(1,t)-\theta(0,t)\). The concurrent-only version uses nuisance functions conditional on \(V_{\tilde a}=1\), whereas the all-controls version replaces the control-arm event and censoring nuisance functions by their pooled-control analogues. This replacement requires Assumption~\ref{ass:pool} and the analogous censoring-pooling condition. The $\drmst$ influence function is obtained by summing over
\(t=1,\ldots,\tau-1\).

\begin{theorem}[Influence function of the fixed-time survival contrast]\label{theorem2}
Let $X=(E,W)$ then under Model \ref{eq:npsem} and assumptions \ref{ass:exchange}-\ref{ass:pconc}  
\begin{enumerate} 
\item The influence function $\phi^{oc}(Z,t)$ for the fixed-time survival contrast is given by: 
\begin{align*}
    &\alpha\times\Big[  S(t\mid A=1, \Va=1, X) -   S(t\mid A=0, \Va=1, X)\Big] \\
    &-\alpha\times \sum_{k=1}^{t} Q_{1}^{oc}(k,Z) + \alpha\times  \sum_{k=1}^{t}Q_{0}^{oc}(k,Z) \\
    &- \alpha\times\Big[(\theta(1,t) - \theta(0,t) )\Big]
\end{align*}
where $\alpha = \dfrac{\I(\Va=1)}{\Prob(\Va=1)}$ and  we define $Q_{1}^{oc}$ as:
\begin{align*}
\dfrac{A\times \I( \tilde{T}>k-1) \times S(t\mid A=1, \Va=1,X)\times \{\I(\tilde{T}=k, \Delta=1)-h(k,A=1,\Va=1,X)\} }{G(k-1\mid A=1,\Va=1,X)S(k\mid A=1,\Va=1,X)\pi(A=1\mid \Va=1,X)}
\end{align*}
similarly we define $Q_{0}^{oc}$ as:
\begin{align*}
\dfrac{(1-A)\times \I( \tilde{T}>k-1) \times S(t\mid A=0, \Va=1, X)\times \{\I(\tilde{T}=k, \Delta=1)-h(k,A=0,\Va=1,X)\} }{G(k-1\mid A=0,\Va=1,X)S(k\mid A=0,\Va=1,X)\pi(A=0\mid \Va=1, X)}
\end{align*}
\item  Under Model~\ref{eq:npsem}, Assumptions~\ref{ass:exchange}--\ref{ass:pool}, and the censoring-pooling condition stated above, the influence function $\phi^{ac}(Z,t)$ for the fixed-time survival contrast is equal to: 
\begin{align*}
    &\alpha\times\Big[  S(t\mid A=1, \Va=1, X) -   S(t\mid A=0,  X)\Big] \\
    &-\alpha\times \sum_{k=1}^{t} Q_{1}^{ac}(k,Z) + \beta\times  \sum_{k=1}^{t}Q_{0}^{ac}(k,Z) \\
    &- \alpha\times\Big[(\theta(1,t) - \theta(0,t) )\Big]
\end{align*}
where $\alpha = \dfrac{\I(\Va=1)}{\Prob(\Va=1)}$, $\beta = \dfrac{\Prob(V_{\tilde a}=1\mid X)}{\Prob(V_{\tilde a}=1)}$ and we define $Q_{1}^{ac}$ as: 
\begin{align*}
\dfrac{A\times \I( \tilde{T}>k-1) \times S(t\mid A=1, \Va=1,X)\times  \{\I(\tilde{T}=k, \Delta=1)-h(k,A=1,\Va=1,X)\} }{G(k-1\mid A=1,\Va=1,X)S(k\mid A=1,\Va=1,X)\pi(A=1\mid \Va=1,X)}
\end{align*}
and $Q_{0}^{ac}$ as:
\begin{align*}
\dfrac{(1-A)\times \I( \tilde{T}>k-1) \times S(t\mid A=0, X)\times  \{\I(\tilde{T}=k, \Delta=1)-h(k,A=0,X)\} }{G(k-1\mid A=0,X)S(k\mid A=0,X)\pi(A=0\mid X)}
\end{align*}
\end{enumerate}
\end{theorem}
\vspace{3mm}

\noindent The proof of this result is located in the Supplementary Material. Note that in the derivation of $\phi^{ac}(Z,t)$,  if $V_{\tilde a}$ is a deterministic function of $E$ then $\Prob(V_{\tilde a}=1\mid X) = \I(V_{\tilde a}=1)$, thus $\beta = \alpha$ and the two fixed-time influence functions coincide.
The influence function for $\drmst$ is obtained by summing the fixed-time influence functions:
\(
\phi(Z,\drmst)=\sum_{t=1}^{\tau-1}\phi(Z,t),
\)
where $\phi(Z,t)$ denotes the influence function in Theorem~\ref{theorem2} for the contrast
$\theta(1,t)-\theta(0,t)$. The IFs derived in Theorem \ref{theorem2} suggest the following estimators,

\begin{align*}
\widehat{\drmst}_{DR}^{oc} 
&= \frac{1}{n} \sum_{i=1}^{n} \sum_{t=1}^{\tau-1} \hat{\alpha}_i
\Bigg[
    \hat{S}(t \mid A_i=1, V_{\tilde a,i}=1, X_i)
    - \hat{S}(t \mid A_i=0, V_{\tilde a,i}=1, X_i) \\
&\qquad
    - \Biggl(
        \sum_{k=1}^{t} \hat{Q}_{1}^{oc}(k, Z_i)
        - \sum_{k=1}^{t} \hat{Q}_{0}^{oc}(k, Z_i)
      \Biggr)
\Bigg], \\
\\
\widehat{\drmst}_{DR}^{ac} 
&= \frac{1}{n} \sum_{i=1}^{n} \sum_{t=1}^{\tau-1}
\Bigg[
\hat{\alpha}_i
\Big\{
    \hat{S}(t \mid A_i=1, V_{\tilde a,i}=1, X_i)
    - \hat{S}(t \mid A_i=0, X_i)
\Big\} \\
&\qquad
-
\Biggl(
    \hat{\alpha}_i \sum_{k=1}^{t} \hat{Q}_{1}^{ac}(k, Z_i)
    - \hat{\beta}_i \sum_{k=1}^{t} \hat{Q}_{0}^{ac}(k, Z_i)
\Biggr)
\Bigg].
\end{align*}
where
\(
\hat p_V=\frac{1}{n}\sum_{i=1}^{n}\mathbbm{1}\{V_{\tilde a,i}=1\},
\hat{\alpha}_i=\frac{\mathbbm{1}\{V_{\tilde a,i}=1\}}{\hat p_V},
\hat{\beta}_i=\frac{\widehat{\Pr}(V_{\tilde a}=1\mid X_i)}{\hat p_V}. \) To compute the desired estimate using $\widehat{\drmst}_{DR}^{oc}$, one would first obtain $\hat{S}$ by learning the hazard function for the event denoted $\hat{h}$, with parametric or machine learning methods on the appropriate conditioning set and then employing the product formulation given in Proposition \ref{prop1}. The quantities denoted $\hat{Q}_{1}^{oc}$ and $\hat{Q}_{0}^{oc}$ are computed by learning the appropriate component functions: $\hat{h}$, $\hat{g}$ and $\hat{\pi}$, using parametric or data adaptive methods (again invoking the appropriate product formulation to obtain $\hat{G}$ and $\hat{S}$) and then plugging in the results according to part one of Theorem \ref{theorem2}. The same process could be repeated to compute $\widehat{\drmst}_{DR}^{ac}$, where $\hat{Q}_{1}^{ac}$ and $\hat{Q}_{0}^{ac}$ are calculated using part two of Theorem \ref{theorem2}.

\subsection{Large sample properties}

The IF in Theorem \ref{theorem2} is identical to that for the treatment specific survival curve in the concurrent population $\Va=1$ and so inherits the same asymptotic properties \citep{west2024}. If the nuisance models required to estimate 
$G$, $S$, and $\pi$ are correctly specified it can be shown that the resulting estimators  $\widehat{\drmst}_{DR}^{oc}$ and $\widehat{\drmst}_{DR}^{ac}$ are root-n consistent, asymptotically normal and have valid $95 \%$ confidence intervals given by  $\widehat{\drmst}_{DR}^{oc} \pm 1.96\sqrt{\widehat{\Var}\{\phi^{oc}(Z,\drmst)\}/n}$ and $\widehat{\drmst}_{DR}^{ac} \pm 1.96\sqrt{\widehat{\Var}\{\phi^{ac}(Z,\drmst)\}/n}$. For $\widehat{\drmst}_{DR}^{ac}$, these properties additionally require Assumption~\ref{ass:pool} and the censoring-pooling condition stated above, so that the pooled control nuisance functions represent the corresponding concurrent-control nuisance functions.  In addition, both estimators admit a form of double robustness in that if either $S$ or $G$ and $\pi$ are estimated consistently then the resulting estimator will be consistent.

\subsection{On efficiency}
\label{sec:efficiency}

Pooling non-concurrent controls can improve precision for the concurrent estimand only when those controls provide valid information about the nuisance functions for the concurrent control arm and the estimator can use that information efficiently. For the event process, the first requirement is Assumption~\ref{ass:pool}, so that the pooled control hazard represents the concurrent control hazard. For estimators that also pool the censoring mechanism, such as the all-controls doubly robust estimator, the analogous condition is also needed for the control censoring hazard. Without these pooling assumptions, the pooled nuisance functions need not represent the concurrent nuisance functions, and pooling should be viewed as a bias--variance tradeoff relative to the concurrent estimand.

\begin{figure}
\centering
\includegraphics[width=6.5in,height=\textheight]{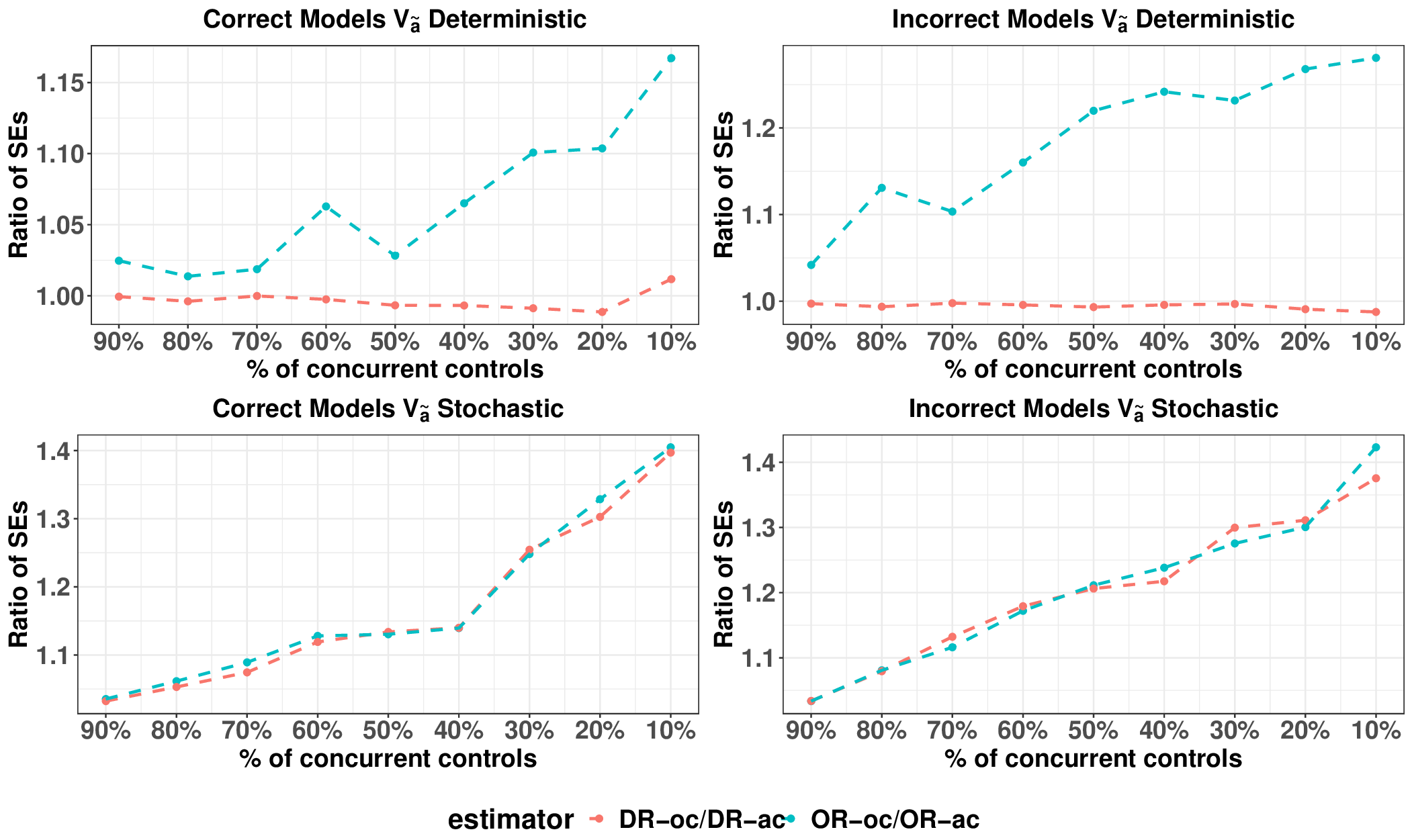}
\caption{Ratio of the estimated standard errors DR-oc/DR-ac and OR-oc/OR-ac across model misspecifications considering $V_{\tilde a}$ as a deterministic function of E (top row) and $V_{\tilde a}$ as a stochastic function of E (bottom row). A ratio greater than 1 indicates a gain in efficiency. }
\label{fig:ratio}
\end{figure}

For parametric outcome-regression estimators, pooling can improve precision through the hazard model. When Assumption~\ref{ass:pool} holds and the fitted hazard model is correctly specified, or otherwise converges to the true concurrent control hazard, non-concurrent controls add observations for estimating the control hazard. This improves estimation of the control hazard, which in turn can improve precision of the plug-in RMST estimator. If either Assumption~\ref{ass:pool} or the hazard-model specification fails, the apparent precision gain may come at the cost of bias. This is seen in the simulations: Figure~\ref{fig:ratio} shows standard-error gains for pooled outcome-regression estimators, while the bias and mean-squared-error panels of Figure~\ref{fig:results2} show that these gains can be outweighed by bias under model misspecification. A more detailed discussion is given in Section~\ref{sec:appendix_A7} of the Supplementary Material.

For doubly robust estimators, the efficiency result follows directly from the influence functions in Theorem~\ref{theorem2}. When treatment availability is deterministic in entry time, $V_{\tilde a}$ is a function of $E$ and hence of $X=(E,W)$. Therefore
\(
\beta
=
\frac{\Prob(V_{\tilde a}=1\mid X)}{\Prob(V_{\tilde a}=1)}
=
\frac{\mathbbm{1}\{V_{\tilde a}=1\}}{\Prob(V_{\tilde a}=1)}
=
\alpha .
\)
Moreover, on the concurrent support, the pooled-control nuisance functions reduce to the corresponding concurrent-control nuisance functions. Thus, in the deterministic-availability setting, the all-controls and concurrent-only influence functions coincide, $\phi^{ac}=\phi^{oc}$, so pooling non-concurrent controls gives no first-order efficiency gain for the doubly robust estimator in the nonparametric model. This is reflected in the top row of Figure~\ref{fig:ratio}, where the standard-error ratio for the concurrent-only and all-controls DR estimators is approximately one.

When treatment availability is stochastic conditional on entry time, the two influence functions need not coincide because $\beta=\Prob(V_{\tilde a}=1\mid X)/\Prob(V_{\tilde a}=1)$ is no longer equal to $\alpha=\mathbbm{1}\{V_{\tilde a}=1\}/\Prob(V_{\tilde a}=1)$. Under the structural model above, $V_{\tilde a}$ depends on $E$ and the trial availability mechanism, so $\Prob(V_{\tilde a}=1\mid X)=\Prob(V_{\tilde a}=1\mid E)$, while the event and censoring nuisance functions remain conditional on $X=(E,W)$.

Under Assumption~\ref{ass:pool} and the analogous censoring condition
\(
g(m,0,\tilde a,e,w)=g(m,0,e,w),
\)
non-concurrent controls can then provide valid information about the control hazard and censoring nuisance functions at covariate values shared with concurrent units.
Heuristically, the all-controls influence function replaces the indicator $\mathbbm{1}\{V_{\tilde a}=1\}$ by its conditional mean $\Prob(V_{\tilde a}=1\mid X)$, a Rao--Blackwellization of the selection weight, which can reduce variance when nuisance functions are consistently estimated. This provides the heuristic for the efficiency gains for the all-controls DR estimator in the bottom row of Figure~\ref{fig:ratio}. Thus, for DR estimation, the practical default is to use concurrent controls only unless Assumption~\ref{ass:pool}, the analogous censoring-pooling condition, and any additional modeling assumptions are credible.

\section{Simulations}
\label{sec:sim}

In this section we evaluate the performance of the proposed estimators using the following metrics: bias squared, variance (sampling variability), mean squared error and coverage of the $95\%$ confidence interval. Specifically, we investigate estimator performance under varying percentages of concurrent controls (for fixed total trial size), with and without model misspecification. The aim of our simulations is to illustrate the large-sample theory developed earlier, not to evaluate the proposed methods under realistic complex conditions.

\subsection{Data generating process}
\ignore{
We generate data from Model \ref{eq:npsem} using a sample size of of $n = 1500$, specifically for each subject $i = 1,\ldots, n$ 
we simulate the following data $500$ times for each point in the grid of percentage of concurrent controls and tmax = 12 discrete time points, according to the following steps: }

We simulate platform trial data with survival outcomes from Model \ref{eq:npsem} for a fixed sample of size $n=1500$ and $K=12$ discrete time points. The data is generated across a varying percentage of concurrent controls denoted $\rho$ such that $\rho \in\{0.1,\ldots,0.9\}$. For each value of $\rho$, $500$ sets of platform trial data are generated. A complete description of the data generating process can be found in the Supplemental Materials (Section \ref{sec:appendix_dgp}). Here, we provide a summary. 

The estimand of interest is the difference in restricted mean survival time between treated and control groups at time point 8, $\drmst(a=1,\tau=8)$. For each of the 500 datasets generated across the grid of concurrent control percentages we evaluate the performance of the methods listed in Table \ref{methods} in Supplementary Material section \ref{sec:additional_figures}.Performance metrics evaluated were bias squared, sampling variability, mean squared error (MSE) and coverage of the $95\%$ confidence interval. We investigate varying the percentage of concurrent controls between $90\%$ and $10\%$ by intervals of $10\%$. To study the effect of model misspecification, for both the outcome and censoring models, the entry time variable $E$ was omitted and in place of the baseline covariate $W$, a new variable $W^* \sim \text{exp}(\lambda=2)$ was used. Parametric regression models were used to estimate the nuisance functions.

\begin{figure} 
\centering
\includegraphics[width=6.5in,height=\textheight]{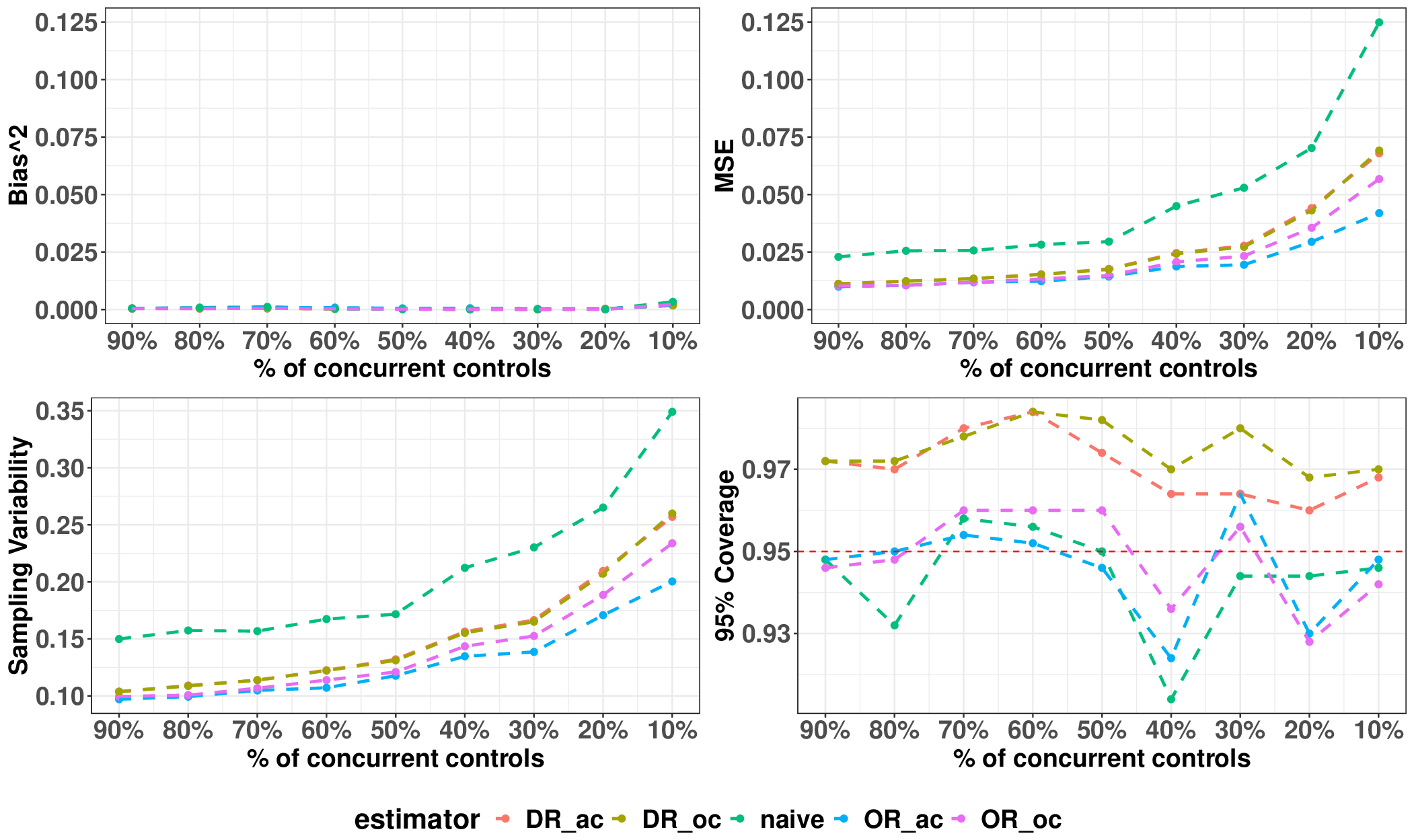}
\caption{Bias squared, mean squared error, variance and coverage of the 95\% confidence interval of the estimators listed in Table \ref{methods} under correct model specification. The DR estimators have substantial overlap in variance and mean squared error.}
\label{fig:results1}
\end{figure}

\subsection{Results}
\label{sec:results}

The simulation results for bias squared, variance, mean squared error, and $95\%$ coverage are shown in Figures \ref{fig:results1} and \ref{fig:results2} in the supplementary. With correctly specified models (Figure \ref{fig:results1}), all methods exhibit negligible bias in estimating $\drmst(a=1,\tau=8)$. As the proportion of concurrent controls decreases, variance and MSE increase for all estimators, with the OR estimator using all controls showing the lowest variance and MSE, indicating potential efficiency gains but relying on the assumptions for pooling concurrent and nonconcurrent controls (Section \ref{sec:appendix_A7}). Coverage remains near nominal for all methods.
Under model misspecification (Figure \ref{fig:results2}), it exhibits increasing bias as the fraction of concurrent controls decreases, leading to inflated MSE despite similar variance across methods. Additional results are shown in the Supplementary Material (Figures~\ref{fig:results2}-\ref{fig:tau12incorrect}). Consequently, its $95\%$ confidence interval coverage falls below nominal. Other estimators, particularly the doubly robust approaches, maintain consistent estimation, correct coverage, and relatively stable variance. In summary, outcome regression with pooled controls can improve efficiency but is sensitive to model misspecification, whereas doubly robust estimators provide reliable performance across scenarios. 

\section{Application to the Adaptive COVID-19 Treatment Platform Trial}
\label{sec:rwd}

\begin{figure}
    \centering
	\includegraphics[width=6.5in,height=\textheight]{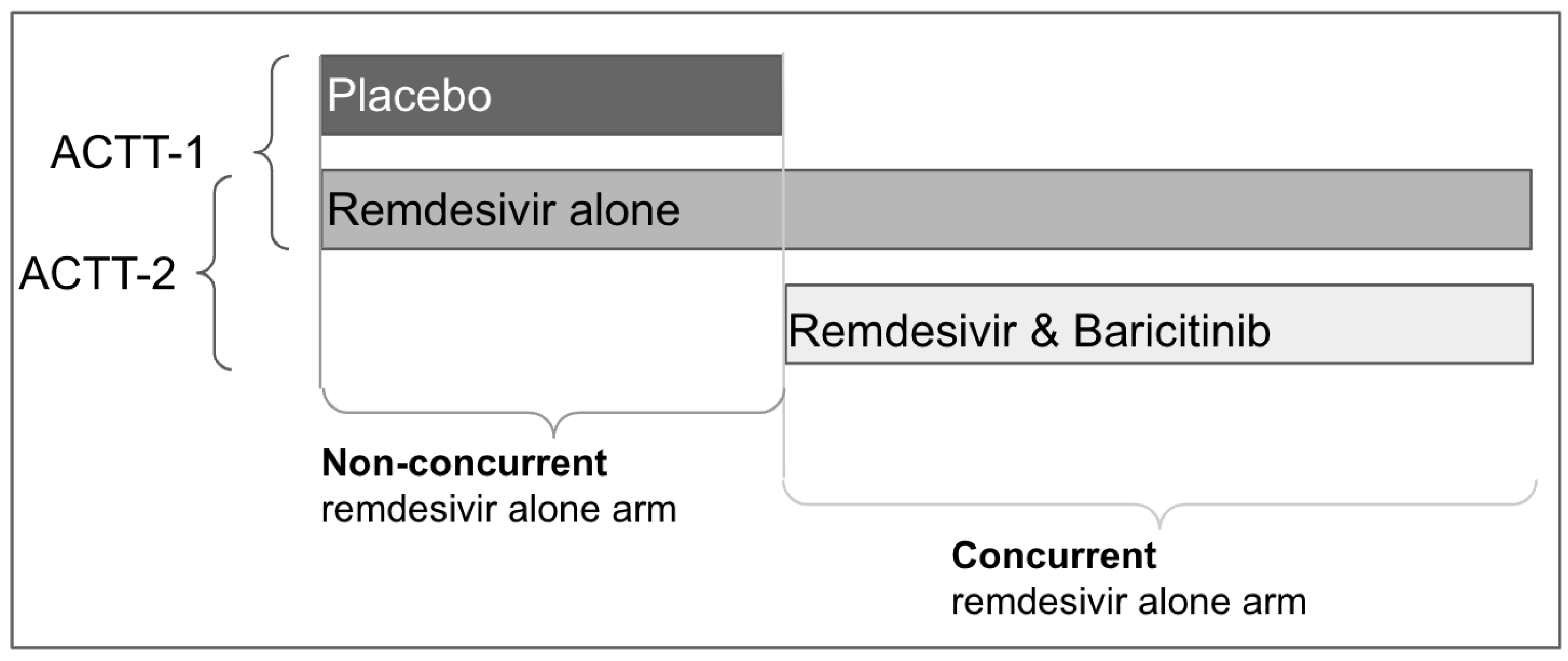}
	\caption{\small Adaptive COVID-19 Treatment Trial (ACTT) schema. Example of concurrent and non-concurrent arm}
	\label{fig:nonconc}
\end{figure}

This manuscript was motivated by the Adaptive COVID-19 Treatment Trial (ACTT) \citep{kalil2021}, a platform trial studying possible treatments for adult patients hospitalized with COVID-19 pneumonia. The trial consisted of multiple stages which are shown in Figure \ref{fig:nonconc}. The first phase, denoted ACTT-1, investigated the effectiveness of remdesivir compared with placebo. In the second phase, denoted ACTT-2, the placebo arm was dropped and a new combined treatment, remdesivir plus baricitinib, was introduced while maintaining remdesivir alone as the \textit{shared arm} (Figure \ref{fig:nonconc}). Patients enrolling during ACTT-2 were randomized to either remdesivir alone or remdesivir plus baricitinib. The data used in this section were accessed using NIAID Clinical Trials Data Repository with the appropriate data user agreement. 

\begin{figure}[h]
    \centering
    \includegraphics[scale=0.45]{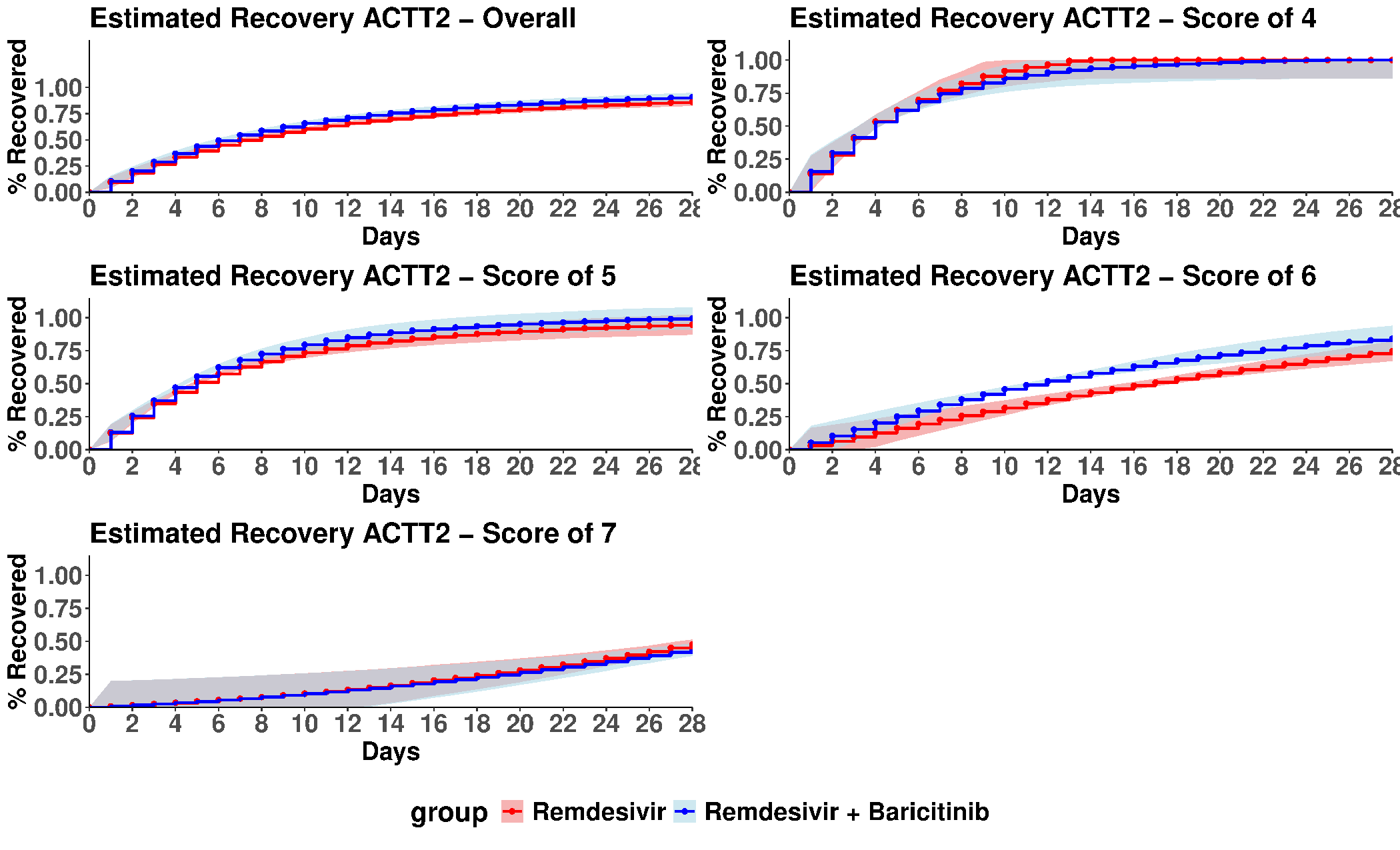}
    
    \caption{Estimated proportion of patients recovered in the combined treatment group, Remdesivir plus Baricitinib compared to the control group Remdesivir alone stratified by disease severity measured at baseline. The estimates were computed using an application of the delta method and \texttt{DR\_oc}.}
    \label{fig:rwdp1}
\end{figure}

\begin{figure}[]
    \centering
   \includegraphics[width=6.5in,height=\textheight]{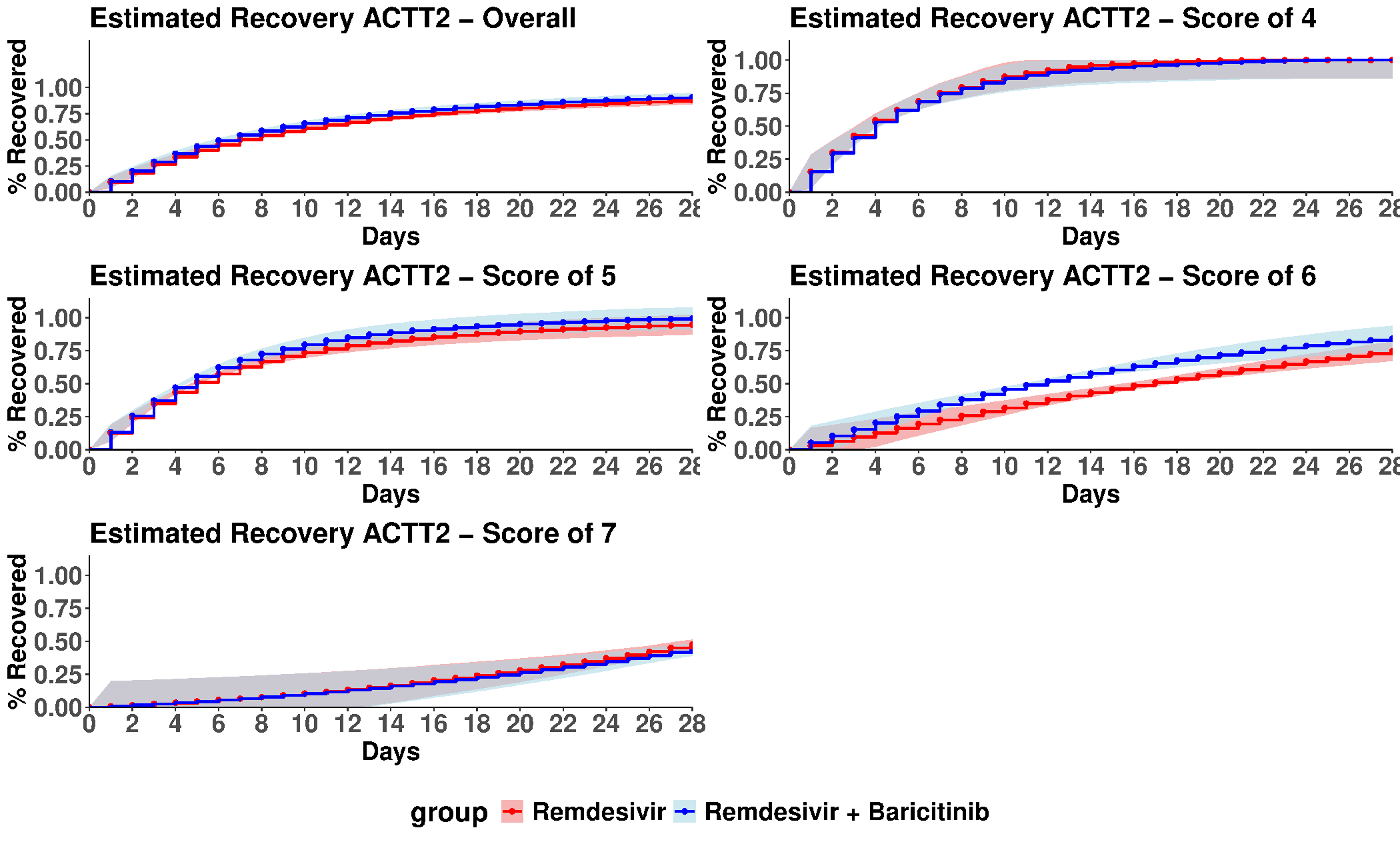} 
   
    \caption{Estimated proportion of patients recovered in the combined treatment group, Remdesivir plus Baricitinib compared to the control group Remdesivir alone stratified by disease severity measured at baseline. The estimates were computed using an application of the delta method and \texttt{DR\_ac}.}
    \label{fig:rwdp2}
\end{figure}

\noindent \textbf{Study population and endpoints.} Our analytic study population was constructed from the combined ACTT-1 and ACTT-2 datasets. We included participants assigned to remdesivir in ACTT-1 ($N=541$), who serve as non-concurrent controls for the ACTT-2 comparison, and participants from ACTT-2 ($N=1033$), who were randomized to remdesivir alone or remdesivir plus baricitinib (Figure \ref{fig:nonconc}). Applying the inclusion and exclusion criteria from the original studies yielded a combined analytic study population of size $N=1379$. The endpoint of interest, following the original trial was time to recovery measured in days.
\textbf{Treatments under study and baseline covariates.} We consider two possible randomized treatment assignments: remdesivir alone (control) and remdesivir plus baricitinib (treatment). The following baseline covariates were considered in our analysis: study entry time, age, sex, body mass index (BMI) and disease severity measured on an ordinal scale. 
\textbf{Estimand and estimation.} 
The causal estimand of interest was a 28-day recovery ratio, motivated by the ``rate ratio for recovery'' reported in the original ACTT-2 analysis \citep{kalil2021}. We define this causal contrast as
\[
\frac{\theta(0,28)}{\theta(\tilde a,28)},
\]
where $\tilde a$ denotes remdesivir plus baricitinib and $0$ denotes remdesivir alone. This estimand was estimated using the methods listed in Table~\ref{methods}. Parametric regression models were used to estimate the nuisance functions. For DR estimators, standard errors were IF-based; for OR estimators, bootstrap-based. For this 28-day recovery ratio, SEs were computed using the delta method, and 95\% Wald confidence intervals and p-values were obtained from Wald tests. Confidence bands for the concurrent treatment-specific survival curves as shown in Figures \ref{fig:rwdp1} and \ref{fig:rwdp2} were computed using a point-wise method as in \cite{west2024}. 
\textbf{Results.} Table \ref{rwdRR} shows the results of the estimators discussed in this work applied to the combined ACTT-1 and ACTT-2 dataset including: point estimate for the 28-day recovery ratio, estimated SE, and Wald $95\%$ CI, p-values and the ratio of the estimated SE compared with that of the naive estimator. Similarly, Table \ref{rwdRMST} contains the same summary data except now estimand of interest is the restricted mean survival time. In addition, Figure \ref{fig:rwdp1} and Figure \ref{fig:rwdp2} display reconstructions, using our methods denoted DR-oc and DR-ac respectively, of the cumulative recovery curves stratified by baseline disease severity. 
Regarding precision, OR and DR estimators using all controls achieved at least a 20\% gain relative to the naive estimator (Table~\ref{rwdRR}). The covariate-adjusted DR estimator using only concurrent data achieved a nearly identical 19\% gain. This suggests that most of the precision improvement comes from covariate adjustment, not from incorporating non-concurrent controls, since the DR estimators using all controls versus only concurrent controls have very similar precision relative to the naive estimator (1.21 vs.\ 1.19).
Regarding sensitivity to pooling and modeling assumptions, in contrast to the continuous-outcome analysis of the same ACTT data by \citet{santa2025}, where the pooled-control OR estimator showed evidence of bias under model misspecification, we do not observe a large divergence across estimators for the 28-day recovery ratio. Still, when the RMST contrast is the target estimand, the OR estimator using all controls shows a slight shift in point estimate relative to the other estimators (Table~\ref{rwdRMST} in the Supplementary Material).
Taken together, these findings indicate that the most robust strategy for improving precision is to target concurrent causal survival estimands and use covariate-adjusted DR estimation with concurrent controls only.

\begin{table}[] 
\centering
\caption{Estimated 28-day recovery ratio using the ACTT data.}\label{rwdRR}
\begin{tabular}{lccccc}
\multicolumn{1}{c}{\textbf{Method}} & $\widehat{\textbf{RR}}$ & \textbf{SE} & \textbf{95\% CI} & \textbf{p-value} & \textbf{Ratio} \\[2pt]
\hline
\textbf{\texttt{OR\_oc}}  & 1.13 & 0.057 & (1.01, 1.23)  & 0.02 & 1.06 \\[2pt]
\textbf{\texttt{OR\_ac}}  & 1.11 & 0.047 & (1.01, 1.20)  & 0.02 & 1.27 \\[2pt]
\textbf{\texttt{DR\_oc}}  & 1.12 & 0.051 & (1.02, 1.22)  & 0.01 & 1.19 \\[2pt]
\textbf{\texttt{DR\_ac}}  & 1.08 & 0.050 & (0.99, 1.18)  & 0.08 & 1.21 \\[2pt]
\textbf{\texttt{naive}}  & 1.12 & 0.061 & (1.00, 1.24) & 0.04 & 1.00 \\[2pt]
\hline
\end{tabular}
\end{table}

\section{Conclusion}
\label{sec:conc}

In this work, we developed an estimand-first causal survival framework for platform trials with time-to-event outcomes. We identified the treatment-specific counterfactual survival curve, \(\theta(a,t)\), in the concurrently eligible population and extended these results to functionals such as \(\drmst\). This framing separates three issues often conflated in analyses using non-concurrent controls: the causal survival estimand, the pooling assumption needed to use non-concurrent controls, and the source of any precision gain. We showed that borrowing non-concurrent controls is not justified by the platform structure alone. It requires a pooling assumption under which the pooled control hazard represents the concurrent control hazard after adjustment for entry time and baseline covariates. Simulations showed that pooled-control outcome-regression estimators can improve precision when this assumption holds and the parametric hazard model is correctly specified, but can induce bias and invalidate inference under model misspecification. For doubly robust estimators, pooling non-concurrent controls gives no first-order efficiency gain when treatment availability is deterministic in entry time, while possible gains under stochastic treatment availability still rely on the required pooling assumptions. These results support the practical recommendation to target concurrent causal survival estimands, avoid pooling unless the required assumptions are credible, and use covariate-adjusted doubly robust estimation with concurrent controls only. Practical guidelines are provided in Section~\ref{sec:prac} of the Supplementary Material.

\section{Disclosure statement}\label{disclosure-statement}

The authors declared no potential conflicts of interest with respect to the
research, authorship, and/or publication of this article.

\section{Funding}

This article is based upon work supported by the National Science Foundation under Grant No 2306556,  and the National Institute of Health Grant No 1R01AI197146-01

\bibliography{bibliography.bib}

\newpage
\phantomsection\label{supplementary-material}
\bigskip

\begin{center}

{\large\bf Supplementary Materials for Causal Survival Analysis in Platform Trials with Non-Concurrent Controls by Antonio D'Alessandro, Samrachana Adhikari and Michele Santacatterina}

\end{center}

\section{A more thorough discussion on the implication of Assumption \ref{ass:pool} for identification and estimation when interested in pooling.}
\label{sec:appendix_A7}

Recall that in this paper we target the concurrent treatment-specific counterfactual survival curve,
\[
\theta(a,t)
:= \Prob\{T(a)>t \mid V_{\tilde a}=1\}
=
\Ex\!\left[\prod_{m=1}^{t}\{1-h(m,a,\tilde a,E,W)\}\ \big|\ V_{\tilde a}=1\right],
\]
for each \(t\in\{0,\ldots,\tau\}\). For the control arm, the concurrent target is therefore
\[
\theta(0,t)
=
\Ex\!\left[\prod_{m=1}^{t}\{1-h(m,0,\tilde a,E,W)\}\ \big|\ V_{\tilde a}=1\right].
\]
When using non-concurrent controls, or ``pooling,'' one instead estimates the control hazard without conditioning on \(V_{\tilde a}=1\),
\[
h(m,0,E,W)
=
\Prob(L_m=1\mid I_m=1,A=0,E,W),
\]
and uses the representation
\[
\theta(0,t)
=
\Ex\!\left[\prod_{m=1}^{t}\{1-h(m,0,E,W)\}\ \big|\ V_{\tilde a}=1\right],
\]
only when Assumption~\ref{ass:pool} justifies replacing \(h(m,0,\tilde a,E,W)\) by \(h(m,0,E,W)\).

Also, recall that treatment availability is known by design and can be deterministic in entry time, i.e. $V_a=\I(E>t)$ for $a\in\{1,\ldots,\mathcal A\}$ and $V_0=1$; or stochastic in entry time, i.e. $V_a=f_{V_a}(E,U_{V_a})$ for $a\in\{0,\ldots,\mathcal A\}$.
We now discuss several questions related to Assumption~\ref{ass:pool}, specifically: 

\begin{enumerate}
    \item Q1: How does assumption \ref{ass:pool} impact \textit{identification} of the concurrent treatment-specific counterfactual survival curve?
    \item Q2: When is Assumption~\ref{ass:pool} plausible?
    \item Q3: How does Assumption~\ref{ass:pool} relate to parametric model specification?
    \item Q4: How does Assumption~\ref{ass:pool} relate to the efficiency of parametric plug-in estimators? That is, how can we formalize the intuition that pooling data increases precision?
\end{enumerate}

\section{Q1: How does Assumption~\ref{ass:pool} impact identification of the concurrent treatment-specific counterfactual survival curve?}

\paragraph*{\(V_{\tilde a}\) stochastic conditional on entry time.}
Suppose treatment availability is stochastic conditional on entry time, so that for some \(e\),
\[
0<\Prob(V_{\tilde a}=1\mid E=e)<1.
\]
We retain \(W\) in the conditioning set because the hazards are defined conditional on \(X=(E,W)\). For values \((m,e,w)\) where the relevant conditional probabilities are well-defined, define
\[
h^{(v)}(m,0,e,w)
=
\Prob(L_m=1\mid I_m=1,A=0,V_{\tilde a}=v,E=e,W=w),
\qquad v\in\{0,1\},
\]
so that \(h^{(1)}(m,0,e,w)=h(m,0,\tilde a,e,w)\). Also define the risk-set mixture weight
\[
p_m(e,w)
=
\Prob(V_{\tilde a}=1\mid I_m=1,A=0,E=e,W=w).
\]
By the law of total probability,
\[
h(m,0,e,w)
=
h^{(1)}(m,0,e,w)p_m(e,w)
+
h^{(0)}(m,0,e,w)\{1-p_m(e,w)\}.
\]
Thus, the pooled hazard is generally a mixture of the concurrent and non-concurrent control hazards. It equals the concurrent control hazard only when this mixture collapses, for example when
\[
h^{(1)}(m,0,e,w)=h^{(0)}(m,0,e,w).
\]
Equivalently, for binary \(L_m\), Assumption~\ref{ass:pool} can be written as
\[
L_m \perp V_{\tilde a}\mid (I_m=1,A=0,E,W),
\]
which implies
\[
h(m,0,e,w)=h(m,0,\tilde a,e,w).
\]
Therefore, when \(V_{\tilde a}\) is stochastic conditional on entry time, Assumption~\ref{ass:pool} is a substantive identification condition for using pooled controls to identify the concurrent control hazard.

\paragraph*{\(V_{\tilde a}\) deterministic in entry time}
Now suppose \(V_{\tilde a}\) is deterministic in entry time, so that \(V_{\tilde a}=v_{\tilde a}(E)\) almost surely. Let
\(
\mathcal E_1=\{e:v_{\tilde a}(e)=1\},\qquad
\mathcal E_0=\{e:v_{\tilde a}(e)=0\}.
\)
For any \(e\in\mathcal E_1\),
\[
\Prob(V_{\tilde a}=1\mid I_m=1,A=0,E=e,W=w)=1,
\]
and hence the mixture weight satisfies \(p_m(e,w)=1\). Therefore,
\[
h(m,0,e,w)=h^{(1)}(m,0,e,w)=h(m,0,\tilde a,e,w),
\qquad e\in\mathcal E_1.
\]
Thus, when availability is deterministic in entry time, Assumption~\ref{ass:pool} holds automatically on the concurrent entry-time support and is not needed to identify the concurrent control hazard there.

\paragraph*{Summary}
Assumption~\ref{ass:pool} matters for identification only when one wants to replace the concurrent control hazard \(h(m,0,\tilde a,e,w)\) by the pooled control hazard \(h(m,0,e,w)\). When \(V_{\tilde a}\) is stochastic conditional on entry time, the pooled hazard is generally a mixture over availability status, so Assumption~\ref{ass:pool} is needed to ensure that pooling does not change the target. When \(V_{\tilde a}\) is deterministic in entry time, this equality holds automatically on the concurrent support. However, in that deterministic case, non-concurrent controls lie outside the concurrent entry-time support; using them to improve precision relies on additional modeling or smoothness assumptions across entry time, not on Assumption~\ref{ass:pool} alone.

\subsection{Q2: When is Assumption~\ref{ass:pool} plausible?}

Assumption~\ref{ass:pool} is a restriction on the true distribution \(\mathsf P_0\). It states that, among control-arm participants who are event-free and uncensored through time \(m\), the event hazard at time \(m\) does not depend on whether treatment \(\tilde a\) was available at entry after conditioning on entry time and baseline covariates, that is, \(h(m,0,e,w)=h(m,0,\tilde a,e,w)\). On strata where both availability levels have positive risk-set probability, this can equivalently be viewed as \(L_m \perp V_{\tilde a}\mid (I_m=1,A=0,E,W)\). Thus, beyond the deterministic-availability case where the equality holds automatically on the concurrent entry-time support, Assumption~\ref{ass:pool} is plausible only when \(V_{\tilde a}\) carries no residual information about the control event process after conditioning on \((E,W)\).
The assumption may be implausible when unmeasured calendar-time, site-level, or practice-pattern factors affect both treatment availability and control outcomes. Let \(U\) denote such factors, for example changes in standard of care, diagnostic intensity, background risk, or enrollment practices. If \(U\) is associated with \(V_{\tilde a}\) within levels of \((E,W)\) and is also prognostic for the event hazard among controls in the risk set, then conditioning on \((E,W)\) may not remove the association between availability and the control event process. In that case,
\(
\Prob(L_m=1\mid I_m=1,A=0,V_{\tilde a}=1,E=e,W=w) 
\neq
\Prob(L_m=1\mid I_m=1,A=0,V_{\tilde a}=0,E=e,W=w),
\)
and therefore the pooled control hazard \(h(m,0,e,w)\) need not equal the concurrent control hazard \(h(m,0,\tilde a,e,w)\).
Conversely, Assumption~\ref{ass:pool} becomes more plausible when entry time \(E\) and measured covariates \(W\) adequately capture the drivers of both treatment availability and baseline prognosis, so that any residual association between \(V_{\tilde a}\) and the control event hazard is negligible.
Assumption~\ref{ass:pool} is an observed-data restriction and, in settings with overlap in treatment availability within levels of \((E,W)\), can be empirically assessed by evaluating whether \(V_{\tilde a}\) predicts the control event hazard among participants in the risk set after adjustment for \((E,W)\). However, such assessments are limited by finite-sample power, modeling choices, and overlap. When \(V_{\tilde a}\) is deterministic in entry time, there is no within-\(E\) variation in availability, so the assumption is not testable as a comparison between concurrent and non-concurrent controls at the same \(E\); on the concurrent entry-time support the equality holds by design, while any use of off-support non-concurrent controls relies on modeling or smoothness assumptions across entry time.

\subsection{Q3: How does assumption \ref{ass:pool} relates to parametric model specification?}

Assumption~\ref{ass:pool} and parametric model specification address different problems. Assumption~\ref{ass:pool} is a restriction on the true distribution \(\mathsf P_0\): it determines whether the pooled control hazard \(h_{\mathsf P_0}(m,0,e,w)\) represents the concurrent control hazard \(h_{\mathsf P_0}(m,0,\tilde a,e,w)\). Thus, it governs \emph{target alignment}. By contrast, parametric model specification determines whether a fitted model consistently estimates whichever hazard it is intended to estimate. Thus, it governs \emph{estimation consistency}.

To make this distinction explicit, let \(\mathcal M=\{\mathsf P_\theta:\theta\in\mathbb R^q\}\) denote a parametric working model for the pooled control hazard and let \(\widehat\theta_{\mathrm{pool}}(0,t)\) denote the corresponding plug-in estimator. If the model is correctly specified, \(\widehat\theta_{\mathrm{pool}}(0,t)\) converges to the pooled control target under \(\mathsf P_0\). If the model is misspecified, it converges instead to the corresponding pseudo-true pooled target induced by the probability limit of the fitted working model. Separately, Assumption~\ref{ass:pool} determines whether this pooled target equals the concurrent control component of the causal estimand.

Table~\ref{tab:a7-model-2x2} summarizes the four resulting cases. The key point is that pooling can fail for two distinct reasons. If Assumption~\ref{ass:pool} fails, then even perfect estimation of the pooled hazard targets the wrong control curve for the concurrent estimand. If the parametric model is misspecified, then even when Assumption~\ref{ass:pool} holds, the fitted pooled hazard may converge to the wrong function. When both fail, pooling combines target mismatch with model bias.

\begin{table}[p]
\centering
\caption{Relationship between Assumption~\ref{ass:pool} and parametric model specification for estimating the concurrent control component \(\theta_{\mathrm{conc}}(0,t;\mathsf P_0)\).}
\label{tab:a7-model-2x2}
\setlength{\tabcolsep}{3pt}
\renewcommand{\arraystretch}{1.12}
\scriptsize
\resizebox{\textwidth}{!}{%
\begin{tabular}{@{}p{0.12\textwidth} p{0.44\textwidth} p{0.44\textwidth}@{}}
\toprule
 & \textbf{$\mathsf P_0 \in\mathcal M$ (well-specified)} & \textbf{$\mathsf P_0\notin\mathcal M$ (misspecified)}\\
\midrule
\textbf{A7 true} &
\textbf{Targeted by pooled estimator:} $\theta_{\mathrm{pool}}(0,t;\mathsf P_0)$, and under A7
$\theta_{\mathrm{pool}}(0,t;\mathsf P_0)=\theta_{\mathrm{conc}}(0,t;\mathsf P_0)$. \par
\textbf{Limit:} $\widehat\theta_{\mathrm{pool}}(0,t)\to \theta_{\mathrm{pool}}(0,t;\mathsf P_0)=\theta_{\mathrm{conc}}(0,t;\mathsf P_0)$. \par
\textbf{Error vs estimand:} none. \par
\textbf{Example (true \& fitted logistic hazard):}
\[
\Prob(L_m{=}1\mid I_m{=}1,A{=}0,E,W,V_{\tilde a}{=}v)
=\expit\{\alpha_m+\beta_m^\top \phi(E,W)\},
\]
(no $v$ term; A7 holds), and the same form is fitted in the pooled controls.
&
\textbf{Targeted by pooled estimator:} $\theta_{\mathrm{pool}}(0,t;\mathsf P_{\theta^\dagger})$, and under A7
$\theta_{\mathrm{pool}}(0,t;\mathsf P_0)=\theta_{\mathrm{conc}}(0,t;\mathsf P_0)$. \par
\textbf{Limit:} $\widehat\theta_{\mathrm{pool}}(0,t)\to \theta_{\mathrm{pool}}(0,t;\mathsf P_{\theta^\dagger})$. \par
\textbf{Error vs estimand:} modeling bias only. \par
\textbf{Example (true logistic, misspecified working model):}
\[
\Prob(L_m{=}1\mid\cdot)=\expit\{\alpha_m+\beta_{1m}E+\beta_{2m}^\top W+\beta_{3m}^\top(E\cdot W)\},
\]
(A7 holds: no $v$ term), but $\expit\{\alpha_m+\beta_{1m}E+\beta_{2m}^\top W\}$ is fitted (omit $E\cdot W$).
\\
\midrule
\textbf{A7 false} &
\textbf{Targeted by pooled estimator:} $\theta_{\mathrm{pool}}(0,t;\mathsf P_0)$. \par
\textbf{Limit:} $\widehat\theta_{\mathrm{pool}}(0,t)\to \theta_{\mathrm{pool}}(0,t;\mathsf P_0)$. \par
\textbf{Error vs estimand:} target mismatch only. \par
\textbf{Example (true logistic violates A7; pooled model correct for pooled hazard):}
\[
\Prob(L_m{=}1\mid\cdot)=\expit\{\alpha_m+\beta_m^\top \phi(E,W)+\gamma_m V_{\tilde a}\},\ \gamma_m\neq 0,
\]
so $h(m,0,e,w)\neq h(m,0,\tilde a,e,w)$, but the pooled conditional mean
$\Prob(L_m{=}1\mid I_m{=}1,A{=}0,E,W)$ is correctly modeled by the working logistic class. 
&
\textbf{Targeted by pooled estimator:} $\theta_{\mathrm{pool}}(0,t;\mathsf P_{\theta^\dagger})$. \par
\textbf{Limit:} $\widehat\theta_{\mathrm{pool}}(0,t)\to \theta_{\mathrm{pool}}(0,t;\mathsf P_{\theta^\dagger})$. \par
\textbf{Error vs estimand:} target mismatch + modeling bias. \par
\textbf{Example (true violates A7 and working model too rigid):}
\[
\Prob(L_m{=}1\mid\cdot)=\expit\{\alpha_m+s_m(E,W)+\gamma_m V_{\tilde a}\},\ \gamma_m\neq 0,
\]
with nonlinear $s_m$, but you fit $\expit\{\alpha_m+\beta_{1m}E+\beta_{2m}^\top W\}$.
\\
\bottomrule
\end{tabular}%
}
\vspace{2pt}
\begin{minipage}{\textwidth}
\scriptsize\noindent\textit{Note on A7 false:} The term \(\gamma_m V_{\tilde a}\) is a reduced-form way to represent residual information about the control hazard carried by treatment availability after conditioning on \((E,W)\). Such residual dependence may arise from unmeasured calendar-time, site-level, or practice-pattern factors \(U\) that affect both availability and the control event process. For example, suppose \(\Prob(L_m{=}1\mid I_m{=}1,A{=}0,E,W,U)=\expit\{\alpha_m+\beta_m^\top\phi(E,W)+\lambda_m U\}\) and \(\Prob(V_{\tilde a}{=}1\mid E,U)=\expit\{\delta_0+\delta_1E+\delta_2U\}\). When \(U\) is both prognostic for the event hazard and associated with treatment availability, conditioning only on \((E,W)\) may leave residual dependence between \(L_m\) and \(V_{\tilde a}\) within the control risk set. The reduced-form coefficient \(\gamma_m\) summarizes this residual association; when \(\gamma_m\neq0\), Assumption~\ref{ass:pool} fails
\end{minipage}
\end{table}
\vspace{-0.5cm}


\normalsize

\subsection{Q4: How does Assumption~\ref{ass:pool} relate to the efficiency of parametric plug-in estimators?}

This subsection clarifies the sense in which pooling can improve precision for parametric plug-in estimators. The key point is that pooling can improve precision only when the additional non-concurrent controls provide valid information about the same control hazard that appears in the concurrent estimand. Assumption~\ref{ass:pool} supplies this target alignment. Parametric model specification supplies estimation consistency.

Fix \(t\in\{1,\ldots,\tau\}\). The concurrent control component is
\[
\theta_{\mathrm{conc}}(0,t;\mathsf P_0)
=
\Ex_{\mathsf P_0}\!\left[
\prod_{m=1}^{t}\{1-h_{\mathrm{conc}}(m,E,W)\}
\mid V_{\tilde a}=1
\right],
\]
where \(h_{\mathrm{conc}}(m,e,w)=\Prob(L_m=1\mid I_m=1,A=0,V_{\tilde a}=1,E=e,W=w)\). Suppose Assumption~\ref{ass:pool} holds, so that \(h_{\mathrm{conc}}(m,e,w)=h_{\mathrm{pool}}(m,e,w)\) on the concurrent domain, and suppose the fitted parametric hazard model is correctly specified, or otherwise converges to this true concurrent control hazard.

Under these conditions, the concurrent-only and pooled-control plug-in estimators target the same survival curve. Their difference is how the control hazard is estimated. The concurrent-only estimator fits the control hazard using control observations with \(V_{\tilde a}=1\), whereas the pooled-control estimator fits the same hazard using the larger control risk set. In a correctly specified parametric likelihood model, this larger risk set increases the information available for the hazard parameters. Thus the fitted pooled-control hazard can be more precise than the fitted concurrent-only hazard. This reduction in hazard-estimation variability then propagates through the product map \(h\mapsto \prod_{m=1}^t\{1-h(m,E,W)\}\) and can reduce the variability of the plug-in estimator of the control survival curve.

This is the sense in which pooling can improve efficiency: it can reduce the hazard-estimation component of the plug-in estimator's asymptotic variance. When Assumption~\ref{ass:pool} holds and the hazard model is correctly specified, non-concurrent controls can improve precision by improving estimation of the same control hazard.

If Assumption~\ref{ass:pool} fails, then \(h_{\mathrm{pool}}(m,e,w)\neq h_{\mathrm{conc}}(m,e,w)\) for some \((m,e,w)\), so the pooled plug-in estimator generally targets a different survival curve than the concurrent estimand. In that case, pooling should be interpreted as a bias--variance tradeoff rather than as an efficiency gain for the concurrent target. If the parametric hazard model is misspecified, pooling may still reduce variance around the pseudo-true fitted hazard, but this does not guarantee smaller bias or mean squared error for the concurrent causal estimand.

The take-away is that pooling can deliver genuine precision gains for parametric plug-in estimators only in the regime where Assumption~\ref{ass:pool} aligns the pooled and concurrent hazards and the fitted hazard model consistently estimates that common hazard. Outside this regime, apparent precision gains should be interpreted cautiously.

\section{Proof of Proposition 1}

Note that Assumption~\ref{ass:pcens}, together with the relevant support conditions, ensures that the censoring survival probabilities appearing below are positive, so the conditional hazards and product representations are well defined.
The survival function may be decomposed as the product of conditional survival probabilities: 

\begin{align*}
    S(m \mid a,1,w,e)&=\Prob(T> m \mid A=a, V_{\tilde a}=1, W=w,E=e)  \\
    &=\prod_{j=1}^{m}\Prob(T>j\mid T>j-1, A=a,V_{\tilde a}=1,W=w,E=e) \\
    &=\prod_{j=1}^{m}1-\Prob(T=j\mid T>j-1, A=a,V_{\tilde a}=1,W=w,E=e) \\ 
\end{align*}
Recall we defined the the hazard function for the event as 
\begin{align*}
 h(m,a,\tilde a,e,w) &= \Prob(L_m=1\mid I_m=1,A=a,V_{\tilde a}=1,E=e,W=w) \\
 &= \Prob(T = m \mid T \geq m, C \geq m, A = a, V_{\tilde a}=1, E=e,  W = w). \notag\\
\end{align*}
Let $X=(E,W)$ and observe that $\Prob(T = m \mid T \geq m, C \geq m, A = a, V_{\tilde a}=1, X=x)$ may be re-written in the following way,  
\begin{align}
    \setcounter{equation}{0}
    &= \dfrac{\Prob(T = m, C \geq m \mid T \geq m, A = a, V_{\tilde a}=1, X=x)}{\Prob(C \geq m \mid T \geq m, A = a, V_{\tilde a}=1, X=x)}  \\
    &= \dfrac{\Prob(T = m \mid T \geq m, A = a, V_{\tilde a}=1, X=x)\Prob(C \geq m \mid A = a, V_{\tilde a}=1, X=x)}{\Prob(C \geq m \mid T \geq m, A = a, V_{\tilde a}=1, X=x)}  \\
    &= \dfrac{\Prob(T = m \mid T \geq m, A = a, V_{\tilde a}=1, X=x)\Prob(C \geq m \mid A = a, V_{\tilde a}=1, X=x)}{\Prob(C \geq m \mid A = a, V_{\tilde a}=1, X=x)}  \\
    &= \Prob(T = m \mid T \geq m, A = a, V_{\tilde a}=1, X=x)
\end{align}
where the first line follows by the definition of conditional probability, line two follows by an application of assumptions \ref{ass:cons} and \ref{ass:cens} to the numerator, line three by applying 
\ref{ass:cons} and \ref{ass:cens} once again to the denominator and line 4 by cancellation of terms. Note here that lines two, three and the simplification in line four are well defined since we assume by the equivalent restatement of Assumption \ref{ass:pcens}: $\Prob(C \geq m \mid A = a, V_{\tilde a}=1, X=x)>0$. Using this application of the definition of conditional probability and identifying assumptions it follows that: 
\begin{align*}
 h(m,a,1,e,w) &= \Prob(T = m \mid T \geq m, A = a, V_{\tilde a}=1, X=x)
\end{align*}
which can be substituted into the product representation of the survival function to yield, 
\begin{align*}
    S(m \mid a,1,w,e)&=\Prob(T> m \mid A=a, V_{\tilde a}=1, W=w,E=e)  \\
    &=\prod_{j=1}^{m}\Prob(T>j\mid T>j-1, A=a,V_{\tilde a}=1,W=w,E=e) \\
    &=\prod_{j=1}^{m}1-\Prob(T=j\mid T>j-1, A=a,V_{\tilde a}=1,W=w,E=e) \\ 
    &=\prod_{j=1}^{m}\{1-h(j,a,\tilde a,e,w)\}.
\end{align*}

Similarly, for the censoring survival function,
\(
G(t\mid A=a,V_{\tilde a}=1,E=e,W=w)
=
\Pr(C\ge t\mid A=a,V_{\tilde a}=1,E=e,W=w).
\)
In discrete time,
\(
G(t\mid A=a,V_{\tilde a}=1,E=e,W=w)
=
\prod_{j=1}^{t-1}
\Pr(C\ge j+1\mid C\ge j,A=a,V_{\tilde a}=1,E=e,W=w).
\)
By the random-censoring assumption, the censoring probability can be evaluated in the event-free risk set:
\(
\Pr(C=j\mid C\ge j,A=a,V_{\tilde a}=1,E=e,W=w)
=
\Pr(C=j\mid T>j,C\ge j,A=a,V_{\tilde a}=1,E=e,W=w)
=
g(j,a,\tilde a,e,w).
\)
Therefore,
\[
G(t\mid A=a,V_{\tilde a}=1,E=e,W=w)
=
\prod_{j=1}^{t-1}\{1-g(j,a,\tilde a,e,w)\}.
\]

\newpage
\section{Proof of Theorem 1}

Fix $t\in\{0,\ldots,\tau\}$ and $a\in\{0,\tilde a\}$. Then
\begin{align*}
\Prob\{T(a)>t\mid V_{\tilde a}=1\}
&= \Ex\!\left[\Prob\{T(a)>t\mid W,E,V_{\tilde a}=1\}\ \big|\ V_{\tilde a}=1\right] \\
&= \Ex\!\left[\Prob\{T(a)>t\mid A=a,W,E,V_{\tilde a}=1\}\ \big|\ V_{\tilde a}=1\right]
&&\text{by Assumption~\ref{ass:exchange}}\\
&= \Ex\!\left[\Prob\{T>t\mid A=a,W,E,V_{\tilde a}=1\}\ \big|\ V_{\tilde a}=1\right]
&&\text{by Assumption~\ref{ass:cons}}\\
&= \Ex\!\left[S(t\mid A=a,V_{\tilde a}=1,W,E)\ \big|\ V_{\tilde a}=1\right]  &&\text{by Def. and \ref{ass:pconc}}\\
&= \Ex\!\left[\prod_{j=1}^{t}\{1-h(j,a,\tilde a,E,W)\}\ \big|\ V_{\tilde a}=1\right]
&&\text{by Proposition~\ref{prop1}}.
\end{align*}

\noindent
For $a=0$, similar arguments yield
\begin{align*}
\Prob\{T(0)>t\mid V_{\tilde a}=1\}
&= \Ex\!\left[\Prob\{T(0)>t\mid W,E,V_{\tilde a}=1\}\ \big|\ V_{\tilde a}=1\right] \\
&= \Ex\!\left[\Prob\{T(0)>t\mid A=0,W,E,V_{\tilde a}=1\}\ \big|\ V_{\tilde a}=1\right]
&&\text{by Assumption~\ref{ass:exchange}}\\
&= \Ex\!\left[\Prob\{T>t\mid A=0,W,E,V_{\tilde a}=1\}\ \big|\ V_{\tilde a}=1\right]
&&\text{by Assumption~\ref{ass:cons}}\\
&= \Ex\!\left[S(t\mid A=0,V_{\tilde a}=1,W,E)\ \big|\ V_{\tilde a}=1\right]
&&\text{by Def. and \ref{ass:pshare}}\\
&= \Ex\!\left[\prod_{j=1}^{t}\{1-h(j,0,\tilde a,E,W)\}\ \big|\ V_{\tilde a}=1\right]
&&\text{by Proposition~\ref{prop1}}\\
&= \Ex\!\left[\prod_{j=1}^{t}\{1-h(j,0,E,W)\}\ \big|\ V_{\tilde a}=1\right]
&&\text{by Assumption~\ref{ass:pool}}.
\end{align*}

\newpage

\section{Proof of Theorem 2}

We assume all data $Z=(X,A,V_a,\Delta,\tilde{T})$ are discrete with a parametric submodel given by, 

\begin{align*}
    \Prob_\epsilon(z) = (1-\epsilon)\Prob_0(z) + \delta_z\epsilon
\end{align*}

\noindent Where $\delta_z$ is the Dirac measure at $Z=z$ and $X=(E,W)$. Given all data are assumed to be discrete the probability mass function is given by:

\begin{align*}
\Prob_{\epsilon}(z) &= (1-\epsilon)\Prob_0(z) + \I(Z=z)\epsilon \\
&= (1-\epsilon)\Prob_0(\tilde{t}, \delta, \atil, 1,x) + \I(\tilde{T} = \tilde{t}, \Delta = \delta, A= \atil,\Va=1,x
)\epsilon
\end{align*}

\noindent To find the influence function we will compute the pathwise derivative with respect to the given parametric submodel as follows:  
\setcounter{equation}{0}

\setcounter{equation}{0}

\begin{align}
\at{\frac{\partial}{\partial\epsilon}\Psi_1}{\epsilon=0}
&=
\at{
\frac{\partial}{\partial\epsilon}
\Ex\Big[
\prod_{m=1}^{t}
\big(1 - h_{\epsilon}(m, A=\atil, \Va=1, X=x)\big)
\mid \Va=1
\Big]
}{\epsilon=0}
\\
&=
\at{
\frac{\partial}{\partial\epsilon}
\Ex\Big[
S_{\epsilon}(t \mid A=\atil, \Va=1, X=x)
\mid \Va=1
\Big]
}{\epsilon=0}
\\
&=
\at{
\frac{\partial}{\partial\epsilon}
\Big[
\sum_{x}
S_{\epsilon}(t \mid A=\atil, \Va=1, X=x)
\Prob_{\epsilon}(x \mid \va=1)
\Big]
}{\epsilon=0}
\\
&=
\sum_{x}
\at{
\frac{\partial}{\partial\epsilon}
S_{\epsilon}(t \mid A=\atil, \Va=1, X=x)
}{\epsilon=0}
\Prob_{0}(x \mid \va=1)
\nonumber
\\
&\quad +
\sum_{x}
S_{0}(t \mid A=\atil, \Va=1, X=x)
\at{
\frac{\partial}{\partial\epsilon}
\Prob_{\epsilon}(x \mid \va=1)
}{\epsilon=0}.
\end{align}

\noindent Consider the sum in equation 4, we may find the required derivative of $\at{\frac{\partial}{\partial\epsilon}\Prob_{\epsilon}(x\mid v_{\tilde a}=1)}{\epsilon=0}$ as  
\begin{align*}
\at{\frac{\partial}{\partial\epsilon}\Prob_{\epsilon}(x\mid v_{\tilde a}=1)}{\epsilon=0} = \frac{\I(\Va=1)}{\Prob_0(\Va=1)}\{\I(X=x)-\Prob_0(x\mid \va=1)\}.
\end{align*}
\noindent Substituting back in to the initial sum and distributing the function $S_0(t\mid A=1, \Va=1,X=x)$ obtain:
\begin{multline*}
   \frac{\I(\Va=1)}{\Prob_0(\Va=1)}\sum_x S_0(t\mid A=\atil,\Va=1,X=x)\I(X=x) \\
   - \frac{\I(\Va=1)}{\Prob_0(\Va=1)}\sum_x S_0(t\mid A=\atil,\Va=1,X=x)\Prob_0(x\mid v_{\tilde a}=1) 
\end{multline*}
\noindent where resulting difference of terms simplifies such that,
\begin{align*}
    \sum_{x}S_0(t\mid A=\atil, \Va=1,x) \at{\frac{\partial}{\partial\epsilon}\Prob_{\epsilon}(x\mid v_{\tilde a}=1)}{\epsilon=0} = \frac{\I(\Va=1)}{\Prob_0(\Va=1)}S(t\mid A=\atil,\Va=1,X) - \frac{\I(\Va=1)}{\Prob_0(\Va=1)}\Psi_1.
\end{align*}
\noindent To complete the derivation it remains to find the derivative in the first term of the sum in equation 4: 
\begin{align*}
\sum_{x}\at{\frac{\partial}{\partial\epsilon}S_{\epsilon}(t\mid A=\atil, \Va=1,X=x)}{\epsilon=0} \Prob_0(x\mid v_{\tilde a}=1). 
\end{align*}
\noindent By definition the survival function may be reformulated using the product representation given by Proposition \ref{prop1} such that,
\begin{multline*}
    \sum_{x}\at{\frac{\partial}{\partial\epsilon}S_{\epsilon}(t\mid A=\atil, \Va=1,x)}{\epsilon=0}  \Prob_0(x\mid v_{\tilde a}=1)
    = \sum_{x}\at{\frac{\partial}{\partial\epsilon} \prod_{m=1}^{t} 1-h_{\epsilon}(m,\atil,1,x) }{\epsilon=0} \Prob_0(x\mid v_{\tilde a}=1)\\
    = \sum_{x}\at{\frac{\partial}{\partial\epsilon} \prod_{m=1}^{t} 1-\Prob_{\epsilon}(L_m \mid I_m=1, A=\atil, \Va=1,X=x) }{\epsilon=0} \Prob_0(x\mid v_{\tilde a}=1)\\
    = \sum_{x}\at{\frac{\partial}{\partial\epsilon} \prod_{m=1}^{t} 1-\Ex_{\epsilon}[L_m \mid I_m=1,A=\atil,\Va=1,X=x] }{\epsilon=0} \Prob_0(x\mid v_{\tilde a}=1).\\
\end{multline*}
\noindent Where the second line follows by definition of the discrete time hazard for the event and the third line by definition of $L_m$ as a binary variable. Using an extension of the product rule from single variable calculus:  
\begin{align*}
\frac{d}{dx}\Big[\prod_{i=1}^{k}f_i(x)\Big] = \sum_{i=1}^{k} \Big[ \Big( \frac{d}{dx}f_i(x) \Big) \prod_{j=1 ; j \neq i}^{k}f_j(x) \Big], 
\end{align*}
we may re-express the sum in the third line,    
\begin{multline*}
     \sum_{x} \Big\{\sum_{k=1}^{t} \Big(\prod_{j\neq k}^{t}1-\Ex_{}[L_j\mid I_j=1,A=\atil,\Va=1,X=x]\Big) \\
     \times  \at{\frac{\partial}{\partial\epsilon} -\Ex_{\epsilon}[L_k \mid I_k=1,A=\atil,\Va=1,X=x] }{\epsilon=0} \Big\}\Prob_0(x\mid v_{\tilde a}=1)
\end{multline*}
which admits further re-expression, 
\begin{multline*}
         = \sum_{x}\Big\{\sum_{k=1}^{t} \dfrac{\prod_{j=1}^{t}1-\Ex[L_j \mid I_j=1, A=\atil, \Va=1, X=x]}{1-h(k,A=a,\Va=1,X=x)} \\
     \times \at{\frac{\partial}{\partial\epsilon} -\Ex_{\epsilon}[L_k \mid I_k=1,A=\atil,\Va=1,X=x] }{\epsilon=0} \Big\}\Prob_0(x\mid v_{\tilde a}=1).
\end{multline*}
We may invoke the product representation of the survival function again to obtain further simplification given by, 
\begin{multline*}
             = \sum_{x}\Big\{\sum_{k=1}^{t} \dfrac{S(t\mid A=\atil,\Va=1,X=x)}{1-h(k,A=\atil,\Va=1,X=x)} \\
     \times \at{\frac{\partial}{\partial\epsilon} -\Ex_{\epsilon}[L_k \mid I_k=1,A=\atil,\Va=1,X=x] }{\epsilon=0} \Big\}\Prob_0(x\mid v_{\tilde a}=1).
\end{multline*}
To continue with the derivation observe, 
\begin{multline*}
    -\at{\frac{\partial}{\partial\epsilon}\Ex_{\epsilon}[L_k \mid I_k=1,A=\atil,\Va=1,X=x] }{\epsilon=0} \\
    = -\sum_{0,1}l_k \times \at{\frac{\partial}{\partial\epsilon} \Prob_{\epsilon}(L_k=1 \mid I_k=1, A=\atil, \Va=1,X=x )}{\epsilon=0}\\
    = -\sum_{0,1}l_k \times \dfrac{\I(I_k=1,A=\atil,\Va=1,X=x)}{\Prob_0(I_k=1\mid A=\atil,\Va=1,X=x)\Prob_0(A=\atil \mid \Va=1,X=x)\Prob_0(x\mid \Va =1)\Prob_0(\Va=1)} \\ 
    \times \Big\{  \I(L_k=1) - \Prob_0(L_k=1\mid I_k=1,A=\atil,\Va=1,X=x  \Big\}.
\end{multline*}
After substitution of the above re-expression, 
\begin{multline*}
    = \sum_{x}\Big\{ \sum_{k=1}^{t}\dfrac{S(t\mid A=\atil,\Va=1,X=x)}{1-h(k,A=\atil,\Va=1,X=x)} \\
    \times \sum_{0,1}l_k \times \dfrac{\I(L_k=1,I_k=1,A=\atil,\Va=1,X=x)}{\Prob_0(I_k=1\mid A=\atil, \Va=1,X=x)\Prob_0(A=\atil \mid \Va=1.X=x)\Prob_0(x\mid \Va=1) \Prob_0(\Va=1)} \\
    - \sum_{0,1}l_k \times \dfrac{\I(I_k=1,A=\atil,\Va=1,X=x)}{\Prob_0(I_k=1\mid A=\atil, \Va=1,X=x)\Prob_0(A=a\mid \Va=1.X=x)\Prob_0(x\mid \Va=1) \Prob_0(\Va=1)} \\
    \times\Prob_0(L_k=1\mid I_k=1, A=\atil, \Va=1, X=x)\Big\} \Prob_0(x\mid \Va=1)
\end{multline*}
The term $\Prob_0(x\mid \Va=1)$ cancels and after further re-arrangement we find, 
\begin{multline*}
    = \sum_{x} \sum_{k=1}^{t} \dfrac{S(t\mid A=\atil,\Va=1,X=x)}{1-h(k,A=\atil,\Va=1,X=x)}\\
    \times\dfrac{\I(I_k=1,A=\atil,\Va=1,X=x)}{\Prob_0(I_k=1\mid A=\atil, \Va=1,X=x)\Prob_0(A=\atil \mid \Va=1.X=x)\Prob_0(\Va=1)} \\
    \times\Big( \sum_{0,1}l_k\I(L_k=1) - \sum_{0,1}l_k\Prob_0(L_k=1\mid I_k=1,A=\atil,\Va=1,X=x)   \Big). 
\end{multline*}
and difference in the parentheses reduces to yield, 
\begin{multline*}
    = \sum_{x} \sum_{k=1}^{t} \dfrac{S(t\mid A=\atil,\Va=1,X=x)}{1-h(k,A=\atil,\Va=1,X=x)}\\
    \times\dfrac{\I(I_k=1,A=\atil,\Va=1,X=x)}{\Prob_0(I_k=1\mid A=\atil, \Va=1,X=x)\Prob_0(A=\atil \mid \Va=1.X=x)\Prob_0(\Va=1)} \\
    \times\Big( \I(L_k=1) - h(k,A=\atil,\Va=1,X=x)   \Big) 
\end{multline*}
Notice taking the sum over all $X=x$ with $\I(X=x)$ reduces the equation further to find, 
\begin{multline*}
    = \dfrac{\I(\Va=1)}{\Prob_0(\Va=1)}\sum_{k=1}^{t} \dfrac{S(t\mid A=\atil,\Va=1,X)}{1-h(k,A=\atil,\Va=1,X)}\\
    \times\dfrac{\I(I_k=1,A=\atil)}{\Prob_0(I_k=1\mid A=\atil, \Va=1,X)\Prob_0(A=\atil \mid \Va=1,X)} \\
    \times\Big( \I(L_k=1) - h(k,A=\atil,\Va=1,X)   \Big). 
\end{multline*}
Recalling the definition of the indicator variable $I_k$ and the product representations for the distribution functions $S$ and $G$ one can re-write the probability in the denominator as 
\begin{multline*}
\Prob_0(I_k=1 \mid A=\atil,\Va=1,X) = G(k-1\mid A=\atil,\Va=1,X) \times S(k-1\mid A=\atil,\Va=1,X)    
\end{multline*}
which after substitution,
\begin{multline*}
    = \dfrac{\I(\Va=1)}{\Prob_0(\Va=1)}\sum_{k=1}^{t} \dfrac{S(t\mid A=\atil,\Va=1,X)}{1-h(k,A=\atil,\Va=1,X)}\\
    \times\dfrac{\I(I_k=1,A=\atil)}{ G(k-1\mid A=\atil,\Va=1,X) \times S(k-1\mid A=\atil,\Va=1,X)  \Prob_0(A=\atil \mid \Va=1,X)} \\
    \times\Big( \I(L_k=1) - h(k,A=\atil,\Va=1,X)   \Big). 
\end{multline*}
Now observe that $ ( 1-h(k,A=\atil,\Va=1,X) ) \times S(k-1\mid A=\atil,\Va=1,X) = S(k\mid A=\atil,\Va=1,X)$ and find, 
\begin{multline*}
    = \dfrac{\I(V_a=1)}{\Prob_0(\Va=1)}\sum_{k=1}^{t} \dfrac{S(t\mid A=\atil,\Va=1,X)}{S(k\mid A=\atil,\Va=1,X)}\\
    \times\dfrac{\I(I_k=1,A=\atil)}{ G(k-1\mid A=\atil,\Va=1,X) \times   \Prob_0(A=\atil \mid \Va=1,X)} \\
    \times\Big( \I(L_k=1) - h(k,A=\atil,\Va=1,X) \Big).
\end{multline*}
Finally, using the definitions of $I_k$, $L_k$, $\tilde{T}$ and $\Delta$ obtain,   
\begin{multline*}
    = \dfrac{\I(\Va=1)}{\Prob_0(\Va=1)}\sum_{k=1}^{t} \dfrac{S(t\mid A=\atil,\Va=1,X)}{S(k\mid A=\atil,\Va=1,X)}\\
    \times\dfrac{\I(\tilde{T}>k-1)\I(A=\atil)}{ G(k-1\mid A=\atil,\Va=1,X) \times   \Prob_0(A=\atil \mid \Va=1,X)} \\
    \times\Big( \I(\tilde{T}=k,\Delta=1) - h(k,A=\atil,\Va=1,X) \Big).
\end{multline*}
Combine this result with the simplification for $\sum_{x}S_0(t\mid A=\atil, \Va=1,X=x) \at{\frac{\partial}{\partial\epsilon}\Prob
_{\epsilon}(x\mid v_{\tilde a}=1)}{\epsilon=0}$ and the desired result follows. The same argument may be used to derive the prospective efficient influence function given in part two of Theorem $\ref{theorem2}$ when $A=0$ by invoking Assumption \ref{ass:pool} to re-express the desired pathwise derivative as, 
\begin{align*}
\at{\frac{\partial}{\partial\epsilon}\Psi_0}{\epsilon=0}
&=
\at{
\frac{\partial}{\partial\epsilon}
\Ex\Big[
\prod_{m=1}^{t}
\big(1 - h_{\epsilon}(m, A=0, \Va=1, X=x)\big)
\mid \Va=1
\Big]
}{\epsilon=0}
\\
&=
\at{
\frac{\partial}{\partial\epsilon}
\Ex\Big[
\prod_{m=1}^{t}
\big(1 - h_{\epsilon}(m, A=0, X=x)\big)
\mid \Va=1
\Big]
}{\epsilon=0}.
\end{align*}
and proceeding along the same lines. 

\section{Construction of outcome regression estimators} \label{sec:appendix_OR}

To construct $\widehat{\drmst}_{OR}^{oc}$ one may use the following procedure:

\begin{enumerate}
    \item Learn $\hat{h}_1(m,A=1,\Va=1,E=e,W=w)$ with standard logistic regression for each discrete time point $m$.
    \item Using $\hat{h}_1$ generate individual hazard predictions for each subject $i$ at each time point $m$.
    \item Construct cumulative product matrix $\widehat{\textbf{S}}_1$ such that each row corresponds to subject $i$ and each column $j$  is equal to $\prod_{m=1}^{j} 1-\hat{h}_1(m,A=1,V_{\tilde a}=1,E=e,W=w)$.
    \item Learn $\hat{h}_0(m,A=0,\Va=1,E=e,W=w)$ using standard logistic regression for each discrete time point $m$. 
    \item Using $\hat{h}_0$, repeat steps 2-3 to find $\widehat{\textbf{S}}_0$.
\end{enumerate}

To construct $\widehat{\drmst}_{OR}^{ac}$ proceed as follows:

\begin{enumerate}
    \item Learn $\hat{h}_1(m,A=1,\Va=1,E=e,W=w)$ with standard logistic regression for each discrete time point $m$.
    \item Using $\hat{h}_1$ generate individual hazard predictions for each subject $i$ at each time point $m$.
    \item Construct cumulative product matrix $\widehat{S}_1$ such that each row corresponds to subject $i$ and each column $j$  is equal to $\prod_{m=1}^{j} 1-\hat{h}_1(m,A=1,V_{\tilde a}=1,E=e,W=w)$.
    \item Learn $\hat{h}_0(m,A=0,E=e,W=w)$ using standard logistic regression for each discrete time point $m$. 
    \item Using $\hat{h}_0$, repeat steps 2-3 to find $\widehat{S}_0$.
\end{enumerate}

\section{Data generating process}\label{sec:appendix_dgp}

A complete description of the steps followed to create the data used in the simulation study can be found below:

\begin{enumerate}
    \item For each subject draw entry time $E_i$ from $E \sim N(0,1)$ and generate baseline covariate $W_i$ as $W_i = -\kappa + 0.8\times E_i + \epsilon_i $ where $\kappa = \frac{1}{n}\sum_{i}^{n}(E_i\times0.8)$ and $\epsilon \sim N(0,1)$.
    \item Create indicator variable $V_{\tilde a,i}$ such that $V_{\tilde a,i} = \I(E_i<b)$ where the constant $b$ is found as the solution to an optimization problem such that $\frac{1}{n}\sum_{i=1}^{n} V_{\tilde a,i} = \rho$.
    \item Generate random treatment assignment $A_i$  such that $A \sim \text{Bernoulli}(1/2)$ when $\Va=1$, otherwise $A_i=0$.
    \item For each subject $i$ and time point $t$, generate the true underlying hazard for the event as  $\text{logit}^{-1}(\eta_i)$ where $\eta_i = -3 + -1.05 \times A_i + 0.2 \times E_i + 1.5 \times W_i + 0.3\times t$. Let $\beta_i$ be the vector of length 12 of event probabilities for subject $i$. 
    \item For each subject $i$ and time point $t$, generate the true underlying hazard for censoring as  $\text{logit}^{-1}(\zeta_i)$ where $\zeta_i = -2.7 +  0.1 \times E_i + 0.15 \times W_i + 0.15\times t$. Let $\sigma_i$ be the vector of length 12 of censoring probabilities for subject $i$. 
    \item For each subject $i$ draw a binary sequence of length 12 from $\text{Binomial}(12,p = \beta_i)$ and let  $T_i$ be the location of the first instance of a $1$ in the sequence. 
    \item For each subject $i$ draw a sequence of length 12 from $\text{Binomial}(12,p = \sigma_i)$ and let the censoring time $C_i$ be the location of the first instance of a $1$ in the sequence.
    \item Construct the observed event time and censoring indicator as $\tilde{T}_i = \text{min}(T_i,C_i)$ and $\Delta_i = \I(T_i<C_i)$ respectively. 
\end{enumerate}

\section{Practical considerations} \label{sec:prac}

\noindent \textbf{What estimands should we target?} Recent regulatory guidance \citep{food2021e9, ich2017} advocates first determining the causal estimand of interest through scientific discussion with subject matter experts before positing a statistical model for the outcome of interest. Only after the estimand and its necessary components (target population, outcome variable, etc.) are established do researchers proceed with selecting an optimal estimation strategy. The organization of this work and the target causal estimand follow the recommendations of \citep{food2021e9}. \\

\noindent \textbf{Integrating the shared control arm.} Following the discussion in Section~\ref{sec:appendix_A7}, we advise against routine pooling of concurrent and nonconcurrent controls unless the required pooling and modeling assumptions are scientifically credible. Combining control subjects requires a sequence of related assumptions which are not testable (random censoring) and are likely to be untenable in real world applications (correct model specification and hazard/censoring function stability over time). Simple numerical studies like those described in Section \ref{sec:sim} show how pooling controls when Assumption \ref{ass:pool} does not hold can invalidate the statistical analysis. \\

\noindent \textbf{Precision via stronger baseline predictors.} Recent work suggests that incorporating stronger predictor variables can improve efficiency in estimating treatment effects \citep{colan2015}. In this article we propose this approach as a means to improve efficiency in place of integrating the complete study population of control subjects. Evidence for this alternative can be found in the real world data application Section \ref{sec:rwd}, specifically the improvement in efficiency obtained by \texttt{OR\_oc} and \texttt{DR\_oc} despite the fact that these estimators only incorporate concurrent control units.  \\

\noindent \textbf{What estimators should we use?} Assuming Model \ref{eq:npsem}, to estimate $\drmst$ we suggest employing the doubly robust estimator using only the concurrent control population denoted \texttt{DR\_oc}. Since the estimator \texttt{DR\_oc} integrates only concurrent controls the previously mentioned difficulties associated with utilizing the entire shared control arm are not a factor in estimation. Researchers do not need to be concerned with the many challenges associated with pooling controls discussed in Section \ref{sec:appendix_A7}, and any estimation bias which may come as a result. Additionally, the estimator maintains the desirable properties discussed in Section \ref{sec:estdr}, i.e. valid statistical inference and additional safeguards against model misspecification compared to outcome regression based techniques. \\

\noindent \textbf{Sample size and power calculations.} Theory and results from the numerical studies in this work suggest an efficiency gain when including non-concurrent control subjects into the data analysis with outcome regression estimators. Although these results appear promising, we advise conducting sample size and power calculations as if only the concurrent control units were available. During the analysis of trial data, if researchers desire efficiency gains, non-concurrent control data may be included as described earlier, but with the understanding that the outcome model must be correctly defined, otherwise treatment effect estimates may be biased and the estimated $95\%$ confidence intervals invalid. During the analysis of trial data, non-concurrent control data may be incorporated in secondary or sensitivity analyses when the required pooling and modeling assumptions are credible, but researchers should recognize that misspecification can bias treatment-effect estimates and invalidate confidence intervals.

\section{Additional tables and figures}\label{sec:additional_figures}

\begin{table}[] 
\centering
\caption{Methods used in the estimation of $\drmst$.}\label{methods}
\begin{tabular}{lc}
\hline
\multicolumn{1}{c}{\textbf{Method}}  & \textbf{Acronym}  \\                                                                         
\hline
DR estimator pooled controls ($\widehat{\drmst}_{DR}^{ac}$, \text{Section } \ref{sec:estdr}) & \texttt{DR\_ac} \\ 

DR estimator only concurrent controls ($\widehat{\drmst}_{DR}^{oc}$, \text{Section } \ref{sec:estdr}) & \texttt{DR\_oc} \\ 

OR estimator pooled controls ($\widehat{\drmst}_{OR}^{ac}$, \text{Section } \ref{sec:estor}) & \texttt{OR\_ac} \\ 

OR estimator only concurrent controls ($\widehat{\drmst}_{OR}^{oc}$, \text{Section } \ref{sec:estor}) & \texttt{OR\_oc} \\ 

OR estimator (time index only) only concurrent controls ($\widehat{\drmst}_{OR}^{oc}$, \text{Section } \ref{sec:estor})     & \texttt{naive}\\ \hline
\end{tabular}
\end{table}

\begin{figure}
\centering
\includegraphics[width=6.5in,height=\textheight]{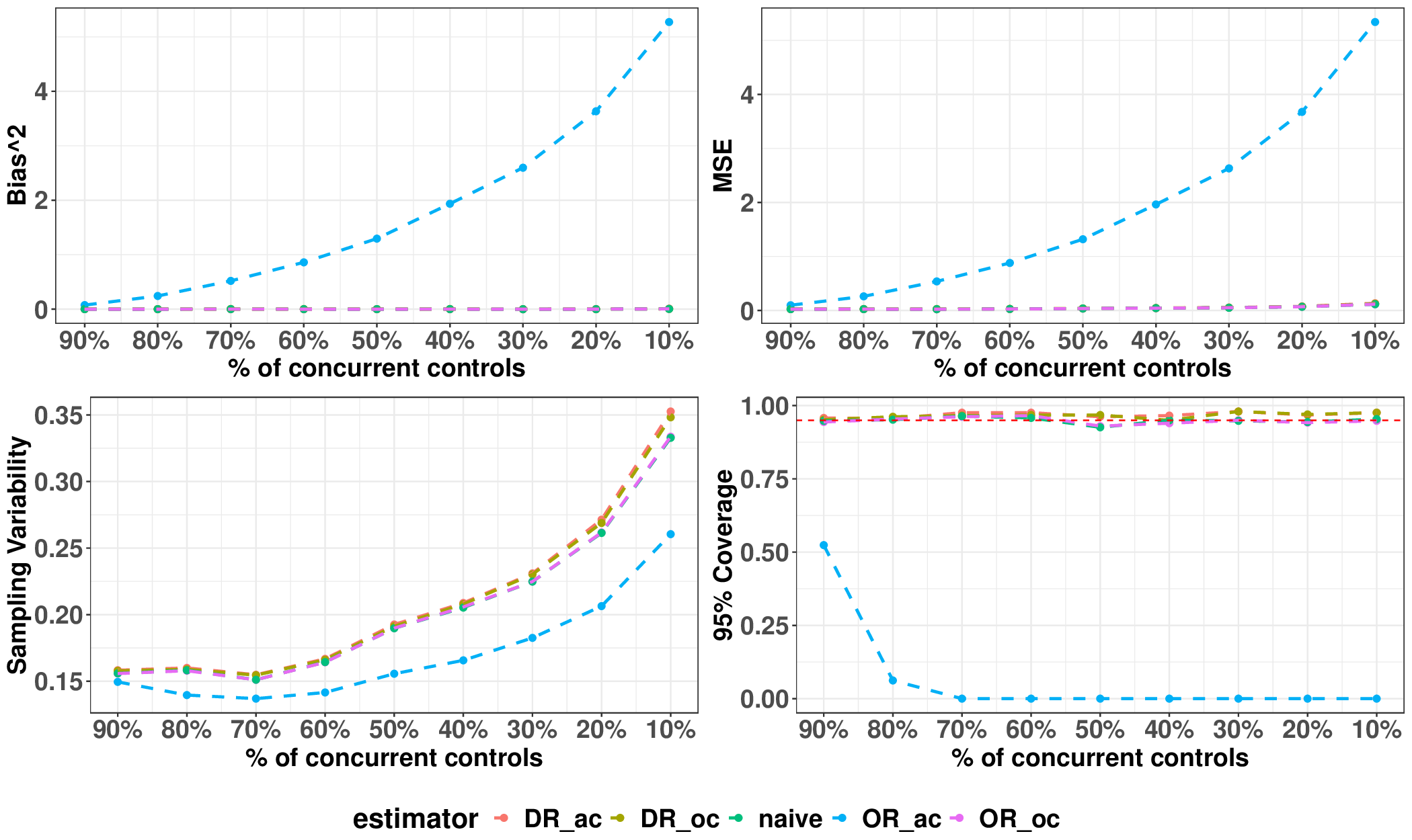}
\caption{Bias squared, mean squared error, variance and coverage of the 95\% confidence interval of the estimators listed in Table \ref{methods} under incorrect model specification. The DR estimators have substantial overlap in variance.}
\label{fig:results2}
\end{figure}

\begin{figure}[h!] 
\centering
\includegraphics[width=6.5in,height=\textheight]{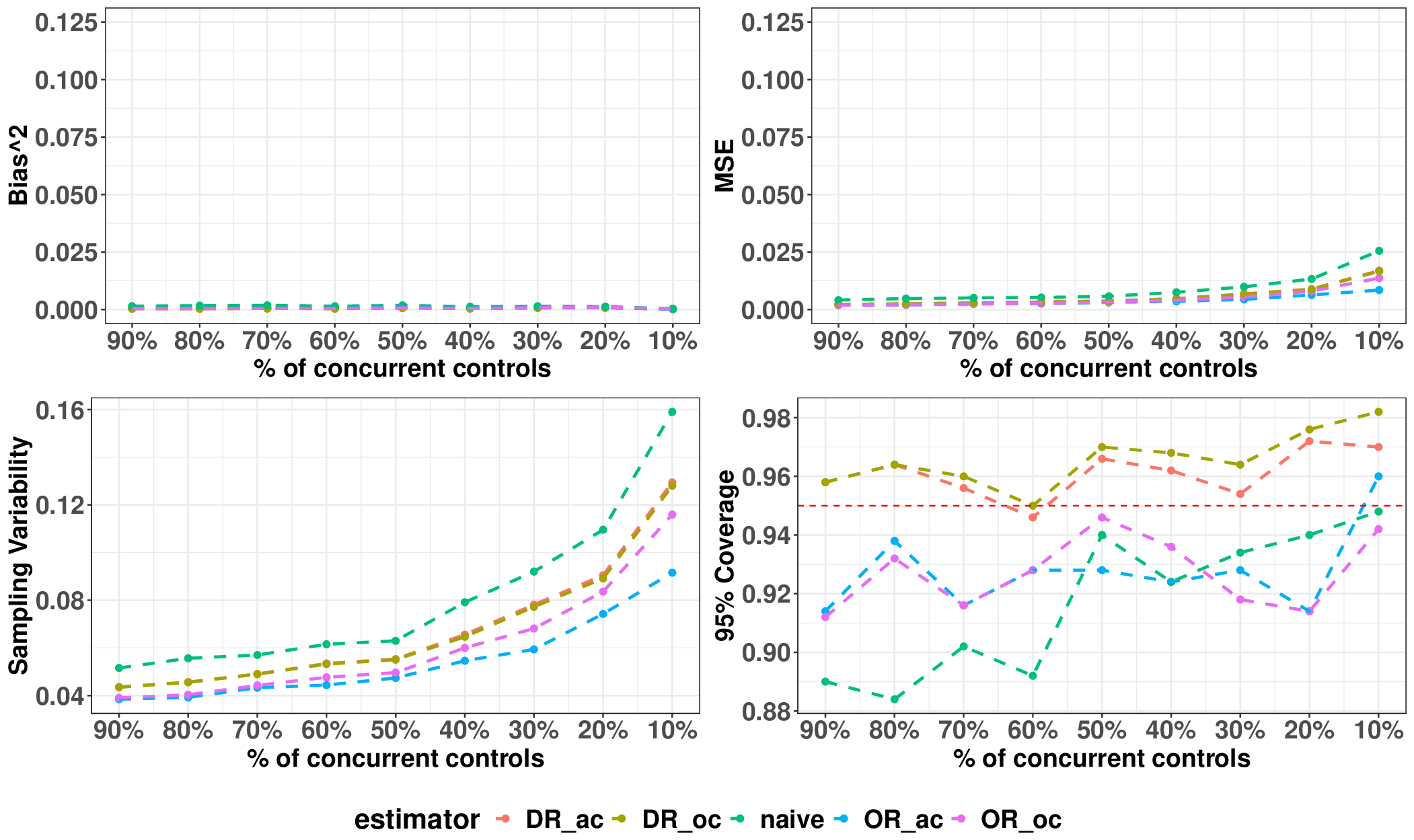}
\caption{Bias squared, mean squared error, variance and coverage of the 95\% confidence interval of the estimators listed in Table \ref{methods} under correct model specification for a discrete time horizon of $\tau=4$. The DR estimators have substantial overlap in variance and mean squared error.}
\label{fig:tau4correct}
\end{figure}

\begin{figure}
\centering
\includegraphics[width=6.5in,height=\textheight]{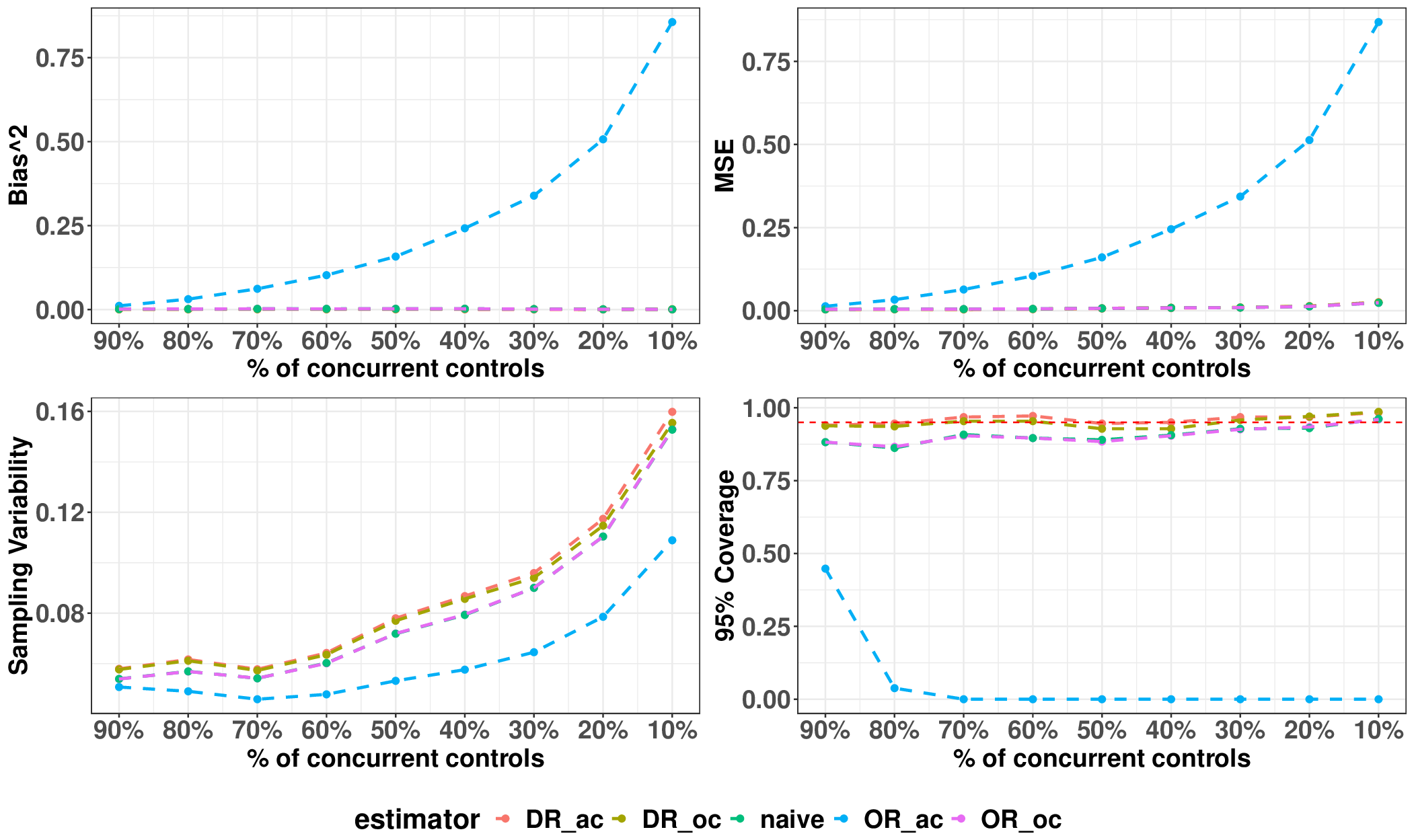}
\caption{Bias squared, mean squared error, variance and coverage of the 95\% confidence interval of the estimators listed in Table \ref{methods} under incorrect model specification for a discrete time horizon of $\tau=4$. The DR estimators have substantial overlap in variance.}
\label{fig:tau4incorrect}
\end{figure}

\begin{figure} 
\centering
\includegraphics[width=6.5in,height=\textheight]{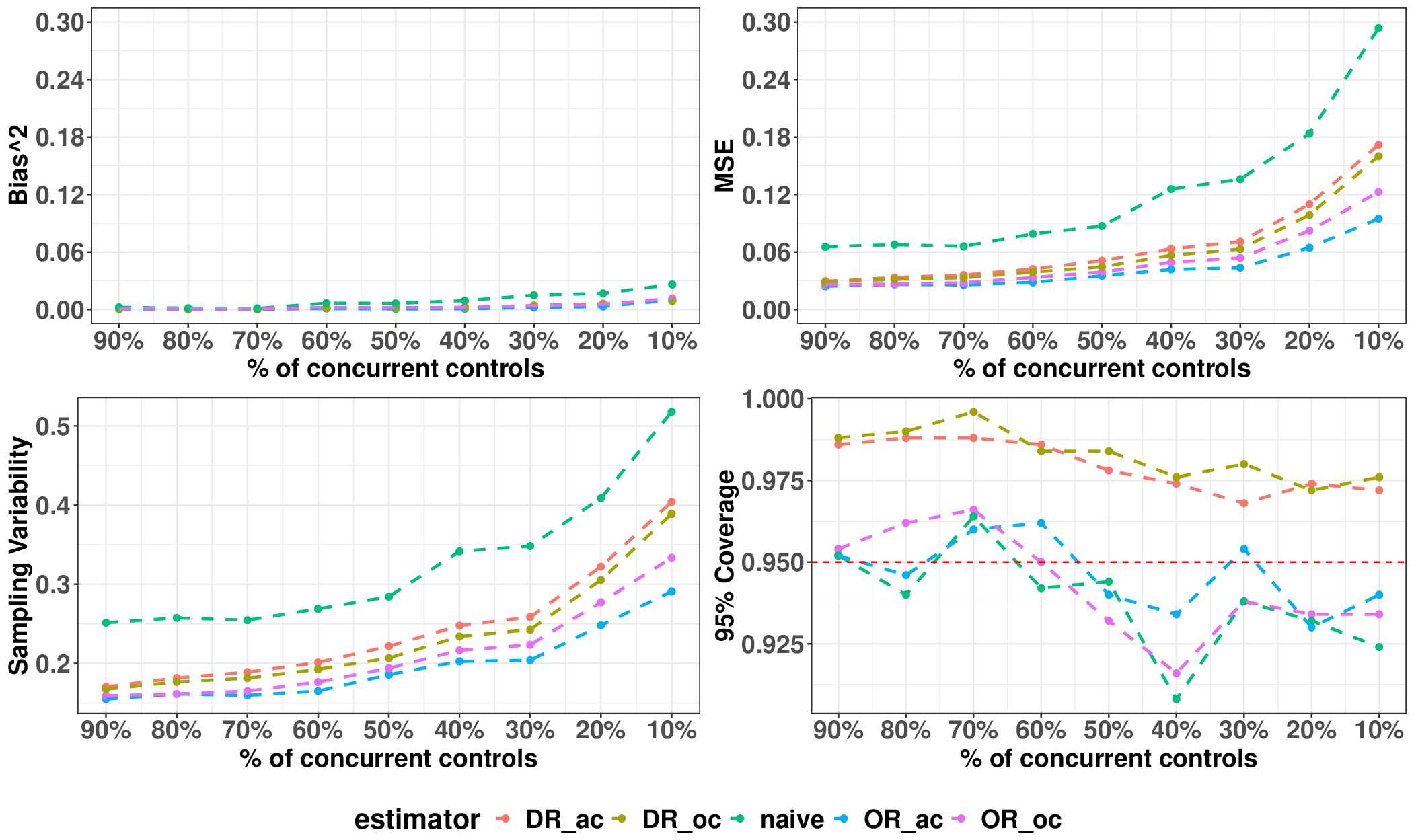}
\caption{Bias squared, mean squared error, variance and coverage of the 95\% confidence interval of the estimators listed in Table \ref{methods} under correct model specification, for a discrete time horizon of $\tau=12$.}
\label{fig:tau12correct}
\end{figure}

\begin{figure}
\centering
\includegraphics[width=6.5in,height=\textheight]{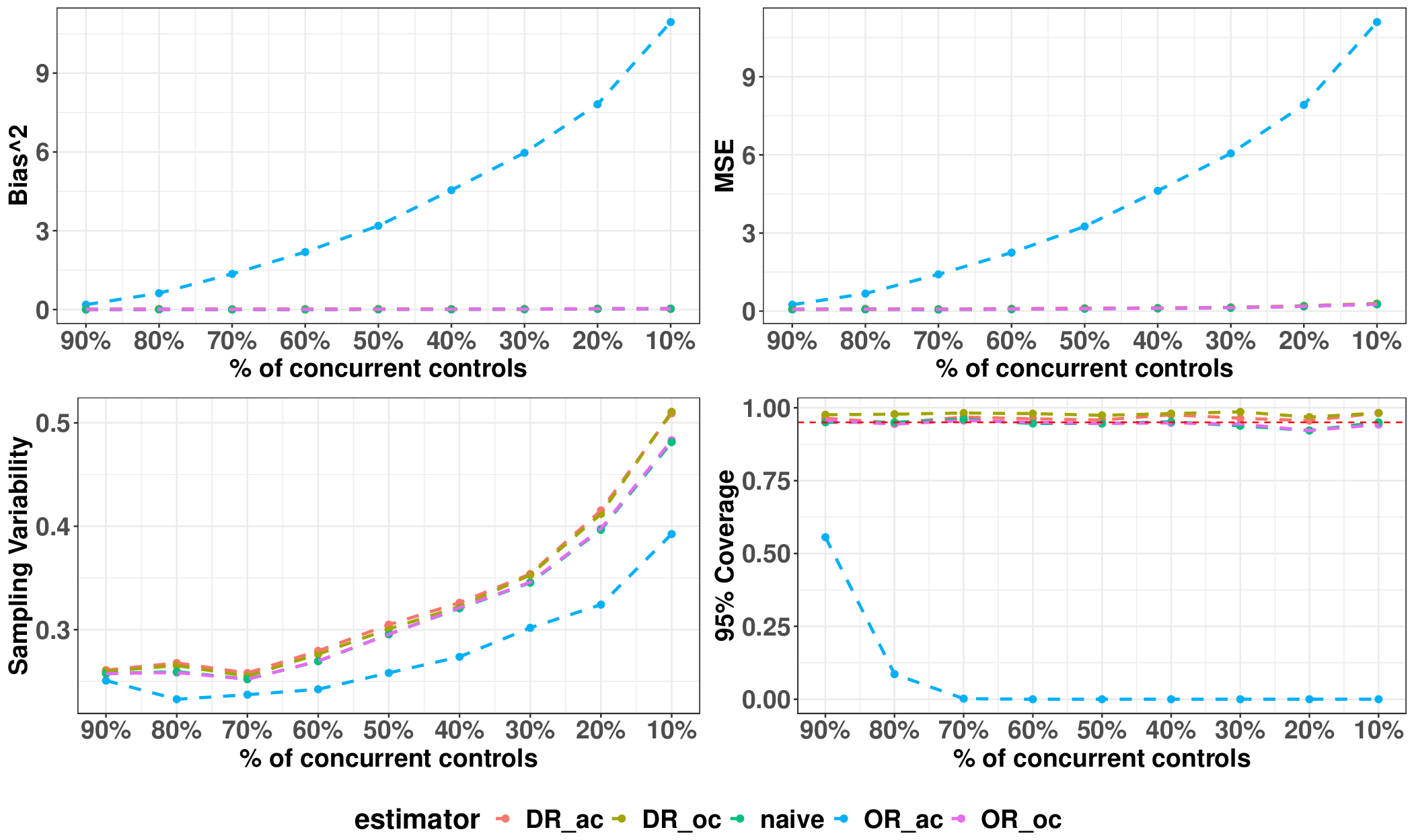}
\caption{Bias squared, mean squared error, variance and coverage of the 95\% confidence interval of the estimators listed in Table \ref{methods} under incorrect model specification, for a discrete time horizon of $\tau=12$. The DR estimators have substantial overlap in variance.}
\label{fig:tau12incorrect}
\end{figure}

\begin{table}[] 
\centering
\caption{Estimated difference in restricted mean recovery time using the ACTT data. Note that following Definition \ref{def:drmst},  $\widehat{\textbf{$\drmst$}}$ is computed between the treatment and control groups, thus a negative estimate corresponds to a faster recovery time among the treated population, in this case an approximate improvement of one day.}\label{rwdRMST}
\begin{tabular}{lccccc}
\multicolumn{1}{c}{\textbf{Method}} & $\widehat{\textbf{$\drmst$}}$ & \textbf{SE} & \textbf{95\% CI} & \textbf{p-value} & \textbf{Ratio} \\[2pt]
\hline
\textbf{\texttt{OR\_oc}}  & -1.21 & 0.46 & (-2.11, -0.30)  & 0.01 & 1.26 \\[2pt]
\textbf{\texttt{OR\_ac}}  & -1.03 & 0.45 & (-1.91, -0.15)  & 0.02 & 1.30 \\[2pt]
\textbf{\texttt{DR\_oc}}  & -1.24 & 0.46 & (-2.14, -0.33)  & 0.01 & 1.26 \\[2pt]
\textbf{\texttt{DR\_ac}}  & -1.24 & 0.46 & (-2.14, -0.33)  & 0.01 & 1.26 \\[2pt]
\textbf{\texttt{naive}}  & -1.22 & 0.58 & (-2.37, -0.08) & 0.04 & 1.00 \\[2pt]
\hline
\end{tabular}
\end{table}

\begin{table}[htbp]
\centering
\caption{Main Estimand for the Adaptive COVID-19 Treatment
Trial}
\renewcommand{\arraystretch}{1.3}
\begin{tabularx}{\textwidth}{>{\bfseries}l X}
\toprule
Objective & To determine the superiority of Remdesivir plus Baricitinib compared with Remdesivir alone in reducing the time to recovery in patients hospitalized adults with Covid-19. \\
\hline
Variable & Time to recovery between baseline and 28 days where recovery is defined as the first day, during the 28 days after enrollment, on which a patient attained category 1, 2, or 3 on an eight-category ordinal scale. \\
\hline
Treatments & Remdesivir plus Baricitinib vs Remdesivir. \\
\hline
Population & Patients hospitalized with COVID-19. \\
\hline
Population-Level Summary & Difference in restricted mean time to recovery through day 28 between treatments. \\
\hline
Intercurrent Events & Post randomization events are not considered.\\ 
\bottomrule
\end{tabularx}
\end{table}

\end{document}